%% file: main.tex
\documentclass[aps,pra,twocolumn,superscriptaddress]{revtex4-2}
\usepackage[utf8]{inputenc}
\usepackage{bm}
\usepackage{bbm}
\usepackage{bbold}
\usepackage{amsfonts}
\usepackage{amsmath}
\usepackage{amssymb}
\usepackage{graphicx}
\usepackage{mathtools}
\usepackage{upgreek}
\usepackage{complexity}
\usepackage{xfrac}
\usepackage{soul}
\usepackage{xcolor}
\usepackage{dsfont}
\usepackage{quantikz}

\newcommand{\sMod}[1]{\ \mathrm{mod}\ #1}

\usepackage[colorlinks=true,citecolor=blue]{hyperref}
\hypersetup{colorlinks=true,citecolor=blue,linkcolor=blue,urlcolor=blue}
\usepackage{url}

\usepackage{txfonts}
\DeclareSymbolFont{matha}{OML}{txmi}{m}{it}
\DeclareMathSymbol{\varv}{\mathord}{matha}{118}

\begin{document}

\title{Advantage for Discrete Variational Quantum Algorithms in Circuit Recompilation}

\author{Oleksandr Kyriienko}
\affiliation{School of Mathematical and Physical Sciences, University of Sheffield, Sheffield S3 7RH, United Kingdom}

\author{Chukwudubem Umeano}
\affiliation{Department of Physics and Astronomy, University of Exeter, Stocker Road, Exeter EX4 4QL, United Kingdom}

\author{Zoë Holmes}
\affiliation{Institute of Physics, \'Ecole Polytechnique F\'{e}d\'{e}rale de Lausanne (EPFL),   Lausanne, Switzerland}
\affiliation{Centre for Quantum Science and Engineering, \'Ecole Polytechnique F\'{e}d\'{e}rale de Lausanne (EPFL),   Lausanne, Switzerland}

\date{\today}

\begin{abstract}
The relative power of quantum algorithms, using an adaptive access to quantum devices, versus classical post-processing methods that rely only on an initial quantum data set, remains the subject of active debate. Here, we present evidence for an exponential separation between adaptive and non-adaptive strategies in a quantum circuit recompilation task. Our construction features compilation problems with loss landscapes for discrete optimization that are unimodal yet non-separable, a structure known in classical optimization to confer exponential advantages to adaptive search. Numerical experiments show that optimization can efficiently uncover hidden circuit structure operating in the regime of volume-law entanglement and high-magic, while non-adaptive approaches are seemingly limited to exhaustive search requiring exponential resources. These results indicate that adaptive access to quantum hardware provides a fundamental advantage.
\end{abstract}

\maketitle

%======================

\section{Introduction}

There are different approaches to combining classical and quantum hardware. Early research in quantum computing focused on algorithms that operated primarily on quantum devices. Over the past decade, much attention has shifted towards variational quantum algorithms (VQAs)~\cite{Cerezo2021rev,bharti2021noisy} and quantum machine learning methods~\cite{wiebe2014quantumdeep, schuld2015introduction, biamonte2017quantum, kyriienko2022protocols} that use hybrid quantum–classical optimization loops. However, the merits of this approach have been called into question, both due to known training challenges~\cite{mcclean2018barren,franca2020limitations,Cerezo2021NatComm,Holmes2022PRXQ, anschuetz2022beyond, Thanasilp2024,Larocca2025rev,crognaletti2025concentration} and to recent results that demonstrate the surprising power of purely classical machine learning algorithms when supplied with quantum-generated data~\cite{Huang2021NatComm, huang2022quantum}. Therefore, the relative strengths and weaknesses of VQAs and quantum-assisted classical methods remain an open question \cite{Schuld2022PRXQ,zimboras2025myths}.

This debate has been framed in several ways. Early work considered the problem in a PAC-learning setting, studying how classical algorithms augmented with quantum data could distinguish phases of matter and learn ground-state properties~\cite{Huang2021NatComm}. In this context, the distinction between the two paradigms mirrors that between the complexity classes BQP and BPP/Samp~\cite{gyurik2023exponential, gilfuster2025traindequantize, jerbi2023shadows}. Later research examined the  task of estimating a loss function at randomly sampled parameter values: does one need repeated quantum evaluations for each parameter value, or can the task be \textit{classically surrogated} in the sense that it can be performed classically after an initial polynomial-time data-collection phase on a quantum device~\cite{basheer2022alternating, jerbi2023power, cerezo2023does, lerch2024efficient}? More recently, heuristic studies ask whether classical algorithms equipped with quantum data can outperform variational quantum algorithms on specific optimization tasks~\cite{bermejo2024qcnns}. Across all these perspectives, the central issue can be viewed as the relative power of adaptive versus non-adaptive access to a quantum device. 

Central to the debate on the relative power of variational quantum algorithms and quantum-assisted classical algorithms is the role of trainability. Variational quantum algorithms are prone to the so-called barren plateau phenomenon, where, for a wide range of parameterized quantum circuits and initialization strategies, loss gradients can be shown to vanish exponentially in system size~\cite{mcclean2018barren, Larocca2025rev}. 
Since quantum systems are subject to finite shot noise, such architectures will generally require exponentially many measurements for training~\cite{Arrasmith_2022, Saem2025Pitfalls}. This prompted the hunt for architectures that provably do not have barren plateaus. However, it was subsequently pointed out that all standard models that provably avoid barren plateaus have loss functions that are classically surrogatable~\cite{cerezo2023does,goh2023liealgebraic,bermejo2024qcnns,rudolph2023classicalsurrogates,rudolph2025pauliprop}. This highlights the need to identify the “Goldilocks zone” of VQAs \cite{Leone2024practicalusefulness,gilfuster2025traindequantize}, where circuits are trainable using realistic resources but are hard to classically surrogate (such that adaptive methods are needed).

In this paper, we provide evidence of an exponential separation between adaptive and non-adaptive methods for the task of discrete circuit compilation. In particular, we construct problems resembling a quantum analog of a hidden bit string version of \texttt{LeadingOnes}, where classically adaptive search methods can provably converge in polynomial time, but non-adaptive strategies require exponential effort \cite{DOERR2019,DOERR2020}. This separation arises from, and holds more generally for, specially structured landscapes that are unimodal (i.e., do not have local minima) and yet non-separable (the parameters cannot be independently optimized)~\cite{DROSTE2002,doerr2018discrete}. Here we translate this principle into a quantum circuit recompilation setting.

Specifically, we consider a circuit recompilation problem, where the objective is to recover hidden T gates (\emph{puzzles}) embedded between layers of partially-random unitaries (Fig.~\ref{fig:sketch}). The algorithm operates on highly-entangled (volume-law) quantum data, where circuits of pre-specified architecture are adaptively adjusted to uncover the locations of the puzzles. We numerically observe an advantage for adaptive optimization strategies over measure-first approaches that grows at increasing system size. We argue that this separation between adaptive and non-adaptive search strategies can be expected, and should persist to larger problem sizes, by identifying a parameter regime where the landscape has non-exponentially vanishing loss differences and yet is approximately unimodal and non-separable. We further argue that the high entanglement and magic of the input states makes this problem robust to known surrogating strategies via classical shadows~\cite{huang2020predicting}. 

Our results can be viewed as providing a concrete example of a discrete variational protocol that navigates the “Goldilocks zone” of being efficiently trainable and yet simultaneously hard to classically surrogate. We note that this claim does not explicitly contradict Ref.~\cite{cerezo2023does} where claims are i) formulated in terms of estimating a loss rather than for the full training procedure, ii) only hold where there is an analytic proof of the absence of barren plateaus and iii) the focus is on continuous variational protocols. Instead we have tackled (heuristically) the issue of when a discrete adaptive method is advantageous for learning.
Crucially, in contrast to prior contrived counterexamples~\cite{cerezo2023does}, our separation holds for a more realistic learning problem without cooking Shor's algorithm (or other cryptographic assumptions) into a loss function~\cite{Shor1997,Liu2021NatPhys,umeano2024simon,Jager2023,umeano2024forrelation}.

% %=====================

\section{Problem Description}

\paragraph*{Problem motivation.} 

Our aim is to find a quantum optimization task that can be efficiently solved by methods that use adaptive access to a quantum device, but cannot be efficiently solved non-adaptively. To do so, we will draw inspiration from a hidden bit-string generalization of the classical optimization problems, namely variants of \texttt{LeadingOnes} and \texttt{OneMax} problems \cite{doerr2018discrete,DROSTE2002}. In the generalized \texttt{LeadingOnes} problem one is given a hidden binary string $x^* \in \{0,1\}^n$ and a fitness function that assigns to each bitstring $x \in \{0,1\}^n$ the number of initial positions in which $x$ and $x^*$ agree. In other words, it counts the length of the longest common prefix of $x$ and $x^*$, stopping at the first index where the two strings differ. The maximum value $n$ is attained uniquely at the target string $x^*$. The landscape corresponding to this loss is \emph{unimodal}, since every bit flip that increases the number of matching bits strictly improves the objective value but \emph{non-separable} since the contribution of each bit cannot be isolated without testing it~\cite{bulanova2022fast}. 

Adaptive algorithms query candidate strings sequentially, where each new query may depend on the outcomes of previous ones. For example, flipping one bit at a time and comparing scores reveals whether that bit matches $x^*$. This allows an adaptive search to identify all $n$ bits in expected $O(n)$ queries. Intuitively, feedback from each query localizes information efficiently, so the hidden string can be reconstructed with polynomial effort. Non-adaptive algorithms, by contrast, must fix all queries in advance. In this case feedback cannot be used to resolve uncertainty bit by bit. To uniquely identify $x^*$, one would require a set of queries that distinguishes all $2^n$ possible strings. 
Thus, while adaptive strategies solve generalized \texttt{LeadingOnes} in polynomial time, non-adaptive approaches require exponentially many queries, establishing an exponential separation between the two regimes~\cite{doerr2018discrete}. 

\begin{figure}[t!]
\includegraphics[width=1.0\linewidth]{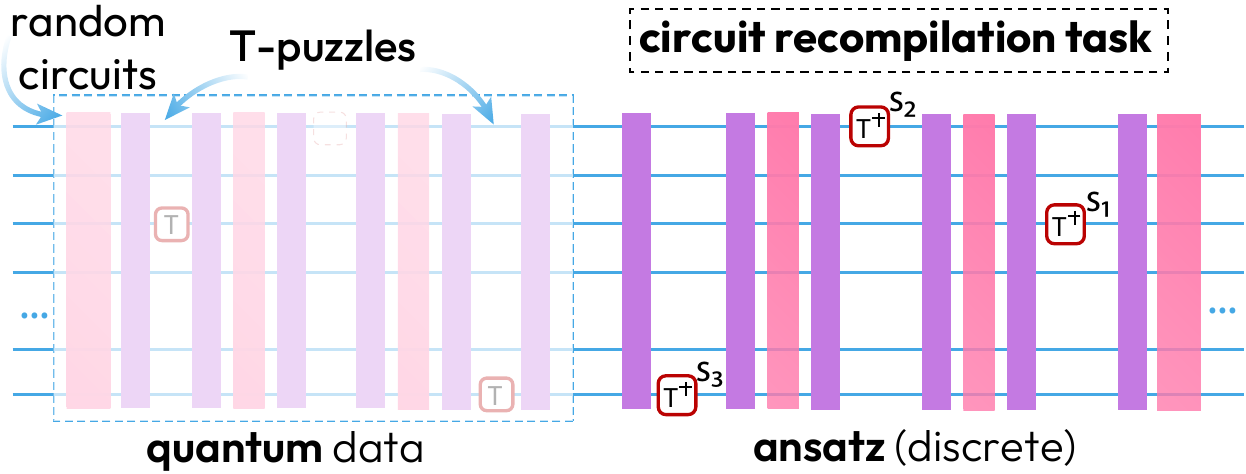}
    \caption{\textbf{Quantum circuit for T gate puzzle task.} Hidden $\mathrm{T}_{q[i]}$ gates are placed on selected qubits $i$ according to a secret bitstring $s = [s_1, s_2, s_3]$, each conjugated by some unitary transformation $V_i$ and layered with circuits $W_i$, representing partially-random unitaries. The positions $q[i]$ are selected at random between $1$ and $n$, and they are known. The resulting quantum state serves as input data for the recompilation task, where the goal is to recover the correct gate placement via discrete optimization. Here we show an example where $q = (3, 1, 6)$ and the solution is $s^*=101$.}
    \label{fig:sketch}
\end{figure}

We propose a quantum circuit recompilation problem, sketched in Fig.~\ref{fig:sketch}, that can be viewed as a quantum analog of generalized \texttt{LeadingOnes}. In \texttt{LeadingOnes}, the task is to identify the hidden binary string $x^*$ by sequentially testing candidate strings. In our setting, the ``hidden bits'' correspond to the positions of T gates (puzzles) embedded between layers of partially random unitaries. The objective is to identify these hidden T gates. Similarly to the \texttt{LeadingOnes} problem, the order in which we identify these strings matters as correctly identifying T gates closer to the target will have a more substantial effect on the loss than later ones. 

More concretely, adaptive strategies can exploit feedback from intermediate recompilation attempts (e.g., correctly identifying early T gates) to gradually 
pinpoint the hidden T gates. However, the importance of the identification order and the interference of the T gates makes the landscape non-separable and so each T gate cannot be learned independently of the others. It follows that non-adaptive strategies seemingly need to explore all candidate recompilations in advance, and thus face a combinatorial explosion of possibilities analogous to the $2^n$ candidate strings in \texttt{LeadingOnes}.
%%%
%%%

In the quantum regime, there are two additional features we need to analyze to understand the relative power of adaptive versus non-adaptive access to quantum hardware: 1) loss concentration and 2) the possibility to surrogate the landscape via classical shadows. We will provide evidence that it is possible to find a sweet spot that balances these additional considerations. Consequently, we expect adaptivity to yield a polynomial-time strategy while non-adaptive 
approaches to require exponential resources.\\

%------------

\paragraph*{Formal problem setup.}

We construct a circuit $U(s^*)$ as a product of interleaved $V_i$-conjugated T-puzzles and partially-random unitaries $W_i$,
\begin{equation}
U(s^*) = \prod_{i=1}^{D} \Big\{ V_i^\dagger (\mathrm{T}_{q[i]})^{s^*_i} V_i \cdot W_i \Big\} \, ,
\label{eq:Us}
\end{equation}
where $s^* = (s^*_1, \dots, s^*_D) \in \mathcal{S} = \{0,1\}^D$ is a bitstring indicating the placement of $D$ non-Clifford gates, $\mathrm{T} = \mathrm{diag}(1, e^{i\pi/4})$, acting on qubit $q[i]$ at the $i$-th layer. The positions $q[i]$ are selected at random between $1$ and $n$, and they are known. The input thus reads as $|\psi_{\mathrm{in}}\rangle = U(s^*) |\mathbf{0}\rangle$, where $|\mathbf{0} \rangle \equiv|0\rangle^{\otimes n}$ is a $n$-qubit register in the computational zero state. We define partially-random unitaries $W, V \propto (1-\beta) \mathds{1} + \beta \mathds{H}$ that interpolate between the identity operator $\mathds{1}$ and a random Hermitian operator $\mathds{H} \equiv \mathds{H}^\dagger$, where $\beta \in \mathbb{R}^+$ tunes the randomness. We choose to use the Cayley transform to enforce the unitarity of $W$ and $V$ \cite{Cayley1846,helfrich2018orthogonalrnns,abanin2025chaos} (see details in Appendix~\ref{sec:Cayley_transform}). 

Next, we define a discrete variational ansatz for solving the problem, provided as a quantum circuit
\begin{equation}
\bar{U}(s) = \prod_{i=D}^{1} \Big\{W_i^\dagger \cdot V_i^\dagger (\mathrm{T}_{q[i]}^\dagger)^{s_i} V_i \Big\},
\label{eq:Ubar}
\end{equation}
where $s \in \mathcal{S}$ is an adjustable bitstring that specifies candidate gate placements out of exponentially many possible answers, $|\mathcal{S}| = 2^D$. Here, circuits for $\mathcal{W} = \{V_i, W_i\}_{i=1}^D$ are given within the problem description, but the challenge is to recover $s^*$.

We use the fidelity between the target $ U(s^*) |\mathbf{0}\rangle$ and trial states $\bar{U}(s) |\psi_{\mathrm{in}}\rangle$ as a simple faithful loss for this problem, 
\begin{equation}
\ell(s) := 1 - | \langle \mathbf{0} | \bar{U}(s) U(s^*) | \mathbf{0} \rangle |^2 \, .
\label{eq:loss_s}
\end{equation}
As an alternative, one can also employ a non-faithful but operationally effective loss based on the variance of a global parity operator (or generally, many-body correlators). In this case, the loss is unity in the highly entangled regime, and drops to zero when we reach a product state (corresponding to the initial state with high certainty); we show the corresponding results for this loss function in Appendix~\ref{sec:diff_loss}. Other more exotic losses could be used to boost sensitivity, with options ranging from covariances for an ensemble of operators~\cite{Boyd2022} to out-of-time-order correlators (OTOC) used for probing quantum chaos \cite{abanin2025chaos}.\\

%========================

\paragraph*{Hill-climbing optimization.} 

To optimize over $s$, we define a fitness function $f(s) := 1- \ell(s)$, which needs to be maximized with discrete optimization, such that $s^{\mathrm{(opt)}} = \text{argmax}_{s \in \mathcal{S}}[f(s)]$. While many discrete optimization algorithms could be applied to this task, we choose hill climbing as the simplest option \cite{doerr2018discrete}. 
Hill climbing is a greedy local search for discrete optimization that works by scanning all single-bit neighbors and flipping the one that yields the smallest loss. More concretely, hill climbing (or rather descent here) proceeds as follows: a) initialize a random bitstring $s^{(0)}$; b) at iteration $m$, compute $\ell\left(s^{(m-1)} \oplus \bm{e}_i\right)$ for all $i=1,\dots,D$, and let $i^*$ be the index returned by $i^* = \operatorname*{arg\,min}_{i} \, \ell\left(s^{(m-1)} \oplus \bm{e}_i\right)$, with corresponding minimal value $\ell_{\min} = \min_i \ell\left(s^{(m-1)} \oplus \bm{e}_i\right)$; c) if $\ell_{\min} < \ell\left(s^{(m-1)}\right)$, update $s^{(m)} = s^{(m-1)} \oplus \bm{e}_{i^*}$, otherwise terminate. This steepest-descent procedure ensures that at each step, the best single-bit improvement is chosen, leading to convergence in landscapes without local traps.

We note that hill climbing is just a simple choice that showcases a minimal adaptivity. More broadly, genetic algorithms could be applied, with population-based updates, potentially balancing exploration and exploitation \cite{MCCALL2005,Katoch2021}. These are expected to work much better on non-trivial landscapes, even though guarantees in this case are typically absent. 
%%%
\begin{figure}[t!]
\includegraphics[width=1.0\linewidth]{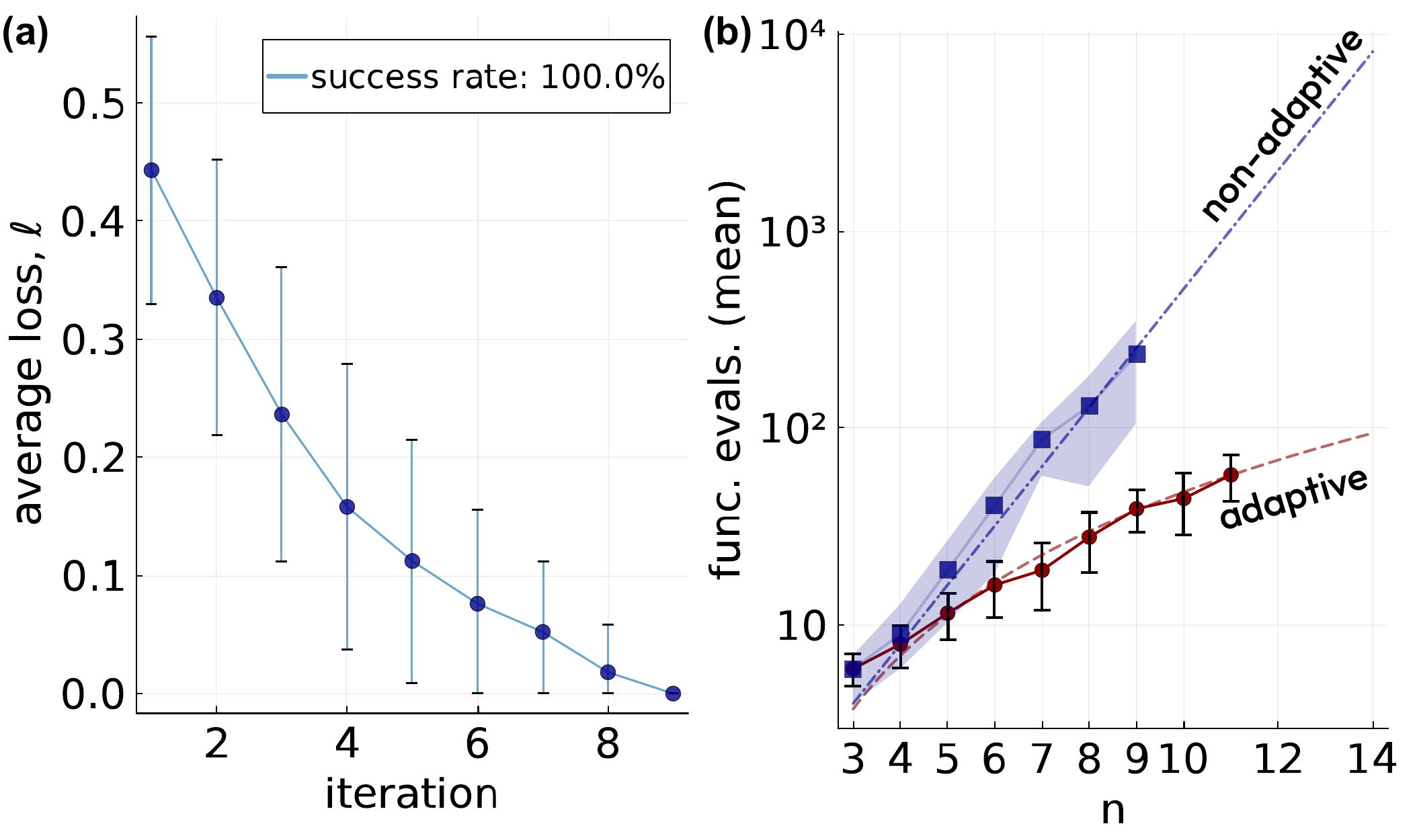}
\caption{\textbf{Convergence and scaling of discrete variational protocol. (a)} Average loss for hill climbing algorithm shows convergence with each iteration, adaptively approaching the global optimum in 50 instances out of 50. The standard deviation is due to different initial bitstrings $s^{(0)}$ and random unitary realizations. We use $n=D=10$ and $\beta_{\mathrm{W}} = \beta_{\mathrm{V}} = 0.2$, but behavior holds for a broad range of parameters. \textbf{(b)} Scaling for the number of function evaluations until convergence at increasing $D=n$, using $50$ instances and logarithmic y-scale. For adaptive search this follows a quadratic scaling, $n^2/2 - n/4$ (red dashed curve), typical of a unimodal landscape. Error bars show a standard deviation. For non-trivial landscapes, non-adaptive methods  are equivalent to random search and follow an exponential scaling, $2^{n}/2$ (blue dot-dashed curve). Shading indicates the interquartile range (25th–75th percentile) across $50$ runs.}
    \label{fig:convergence}
\end{figure}
%%%

%----------------------------

\section{Problem Analysis}

Let us now demonstrate how the described procedure works in practice by performing numerical simulations. First, we set up a puzzle with $D=10$ T gates placed on different $n=10$ qubit wires (shuffled $\{1,\ldots,n\}$), and assign $s^* = 1011010111$. We synthesize partially-random unitaries $\{W_i\}_{i=1}^{10}$ from random Hermitian operators $\mathds{H}$ that are composed of $k=4n^2$ Pauli terms and set the randomness interpolation parameters to $\beta_{\mathrm{W}} = \beta_{\mathrm{V}} = 0.2$. In total $50$ realizations of $\mathcal{W}$ and initial states $s^{(0)}$ are tested, with the hill climbing procedure applied as described above. 

Results are shown in Fig.~\ref{fig:convergence}(a), where each iteration is a sweep round of hill climbing (bit flips).  For all tested instances, the average loss $\ell$ steadily decreases to zero after $D$ iterations, while standard deviations show that optimal solutions can sometimes be achieved after $D/2$ iterations. 
We note that the initialization can be at most a Hamming weight distance $D$ from the solution and for a random initialization the algorithm will on average start $D/2$ away. Therefore, the fact
that the solution can be found after $D$ iterations in all cases (and often after $D/2$ iterations) is indicative of the fact that each iteration of the algorithm brings us one unit step closer to the solution (similarly to hill climbing applied to \texttt{LeadingOnes}).

We continue to test hill climbing for increasing system sizes (both width $n$ and problem size $D$). While in general $n$ and $D$ can be different, we set $n=D$ as a representative combination, and average over $50$ puzzle realizations and different sets $\mathcal{W}$. In Fig.~\ref{fig:convergence}(b) we plot the average number of function evaluations $f_{\mathrm{evals}}$ (i.e. measured loss $\ell$) until convergence to the optimal solution where $\ell(s_{\mathrm{final}}) < 10^{-10}$. Using hill climbing (being a vanilla discrete optimization) we observe a quadratic scaling in the problem size (this follows from the fact that the optimization takes $\Theta(D)$ iterations, each of which require $D$ function evaluations). 
Thus we get advantage from the adaptivity of the protocol (taking informed decisions at each iteration to lower the loss) and the favorable landscape (each iteration is a significant step towards the solution). 
In contrast, a non-adaptive random search on average hits the solution after $f_{\mathrm{evals}} = O(2^n)$ function evaluations.

This is suggestive of a separation between adaptive and non-adaptive approaches to the designed circuit recompilation problem. However, there are a number of issues that still need to be addressed, including: a) its robustness to attacks from predictability, b) whether the loss landscape exhibits exponential concentration and its corresponding robustness to shot noise, c) its robustness to attacks via surrogation using classical shadows and d) its scalability. We tackle these issues in the following subsections. \\

%%%

%===========================

\paragraph{When is the landscape unimodal but non-monotonic and non-separable?} So far we have shown that certain instances of the described recompilation problem can be addressed efficiently with discrete optimization. 
Our next step is to show that variational quantum circuits remain trainable for \emph{typical} instances, while there is no attack on the problem coming from predictability. For this, let us study properties of landscapes, which play an important role in discrete optimization. The objective function $\ell(s)$ over $D$-bit binary strings can be characterized by the following key landscape properties: unimodality, monotonicity, and separability (linearity) \cite{doerr2018discrete}.
%%%
\begin{description}
    \item[Unimodal] A landscape is \emph{unimodal} when there exists a unique global optimum $s^*$, and every other point $s \ne s^*$ has a strictly improving one-flip neighbor $s' ~\text{s.t.}~h(s, s') = 1~\&~f(s') < f(s)$ \cite{DROSTE2002,doerr2018discrete}. Here, $h$ represents the Hamming distance between the two bitstrings. This ensures that adaptive local search methods like hill climbing converge to the global minimum without getting trapped in local minima. 

    We note that unimodal landscapes have relatives in continuous optimization, corresponding to convex landscapes over some weights $\bm{\theta}$. However, in the convex case (generally, a stronger condition), the optimum can be characterized by a system of equations such as $\nabla \ell(\bm{\theta}) = 0$, which can be solved directly if the functional form is known. In the discrete case, however, optimization must typically proceed adaptively, unless the monotonic structure of the landscape is known in advance (analogous to knowing the sign of the gradient in the continuous setting). 

    \item[Monotonicity] refers to functions where moving bitwise in a fixed direction relative to a reference (e.g. optimum $s^*$) consistently improves or worsens the objective. For example, $\ell$ is non-decreasing if $s \leq s'$ (i.e., $s_i \leq s_i'$ for all $i$) implies $\ell(s') \geq \ell(s)$, ensuring predictable behavior as bits change from $0$ to $1$ in a minimization setting. 

    \item[Separable] A loss function is additively \emph{separable} if $\ell(x) = \sum_{i=1}^n \ell_i(x_i)$ with each $\ell_i$ depending only on the $i$-th bit, allowing independent minimization over each bit --- ultimately, an attack on adaptivity as $D$ flips suffice for recompilation in this case.

    A separable discrete loss has analogous separability in the continuous case that is prone to linear sweep of protocols like \texttt{Rotosolve} \cite{Ostaszewski2021structure}.
\end{description}
%%%

Thus we expect a separation between adaptive and non-adaptive methods for landscapes that are unimodal but non-separable and non-monotonic. Below we show that a wide range of parameter instances can lead to such landscapes.  
%%%
\begin{figure}[t!]
\includegraphics[width=1.0\linewidth]{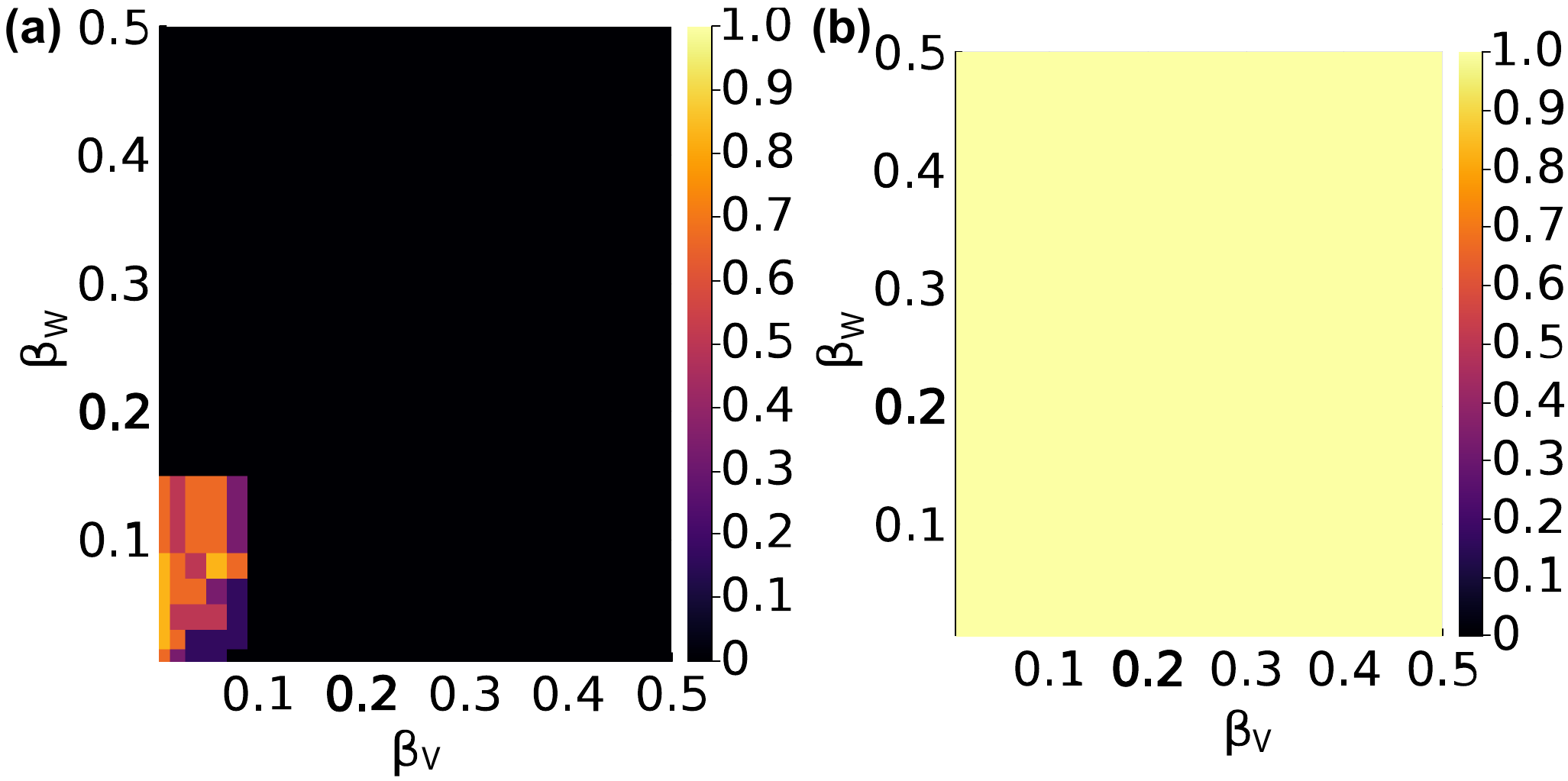}
    \caption{\textbf{Landscape properties of discrete variational circuit recompilation. (a)} Heatmap plot of the fraction of puzzles that are \textit{unimodal} (0 = all unimodal, 1 = all non-unimodal), shown over six problem instances for $n=D=8$, plotted for different $\beta_{\mathrm{W}} = \beta_{\mathrm{V}}$. The region with significant randomness ($\beta_{\mathrm{W}} = \beta_{\mathrm{V}}>0.15$) shows clear unimodality. \textbf{(b)} Heatmap plot of the fraction of problems that are \textit{non-separable} (0 = all separable, 1 = all non-separable), shown for the same instances and system sizes as in (a). The landscape remains non-separable for all problems as there are always non-trivial configurations that prevent the separation. As all randomly selected problem instances were also \textit{non-monotonic} (b) also covers this case.}
    \label{fig:landscape}
\end{figure}
%%%

In Fig.~\ref{fig:landscape} we test the average unimodality, monotonicity and separability for many instances and different $\beta_{\mathrm{W}}$ and $\beta_{\mathrm{V}}$. The values for color bars are calculated as a fraction of solutions that are non-unimodal and non-separable. We see that while for very small $\beta$ the interference leads to non-unimodality, in the range of $\beta_{\mathrm{W}} = \beta_{\mathrm{V}} > 0.15$ the loss is unimodal in all problems studied. We did not observe strictly monotonic or separable instances within the studied range. Namely, there are bitstrings of smaller Hamming weights that are further from optimum than those with larger $h$ (hence, non-monotonic) and different T gates interfere (hence, non-separable). However, this may change with the density of puzzles, as multiple overlapping dressed T gates interfere, so a safe regime to work in is $n \sim D$. \\

%%%

\paragraph{What about exponential concentration and shot noise?} 
Quantum losses, unlike standard classical losses, are unavoidably subject to shot noise, which in effect blurs loss differences. It follows that to effectively train quantum loss landscapes, they need to be sufficiently featured. Or, at the very least, for the algorithm to scale to interesting problem sizes $n$, we require that loss differences do not shrink exponentially fast in $n$. We address this issue in this section. In particular, we show that for a range of moderate $\beta$ the loss differences are insensitive to system size.

\begin{figure}[t!]
\includegraphics[width=1.0\linewidth]{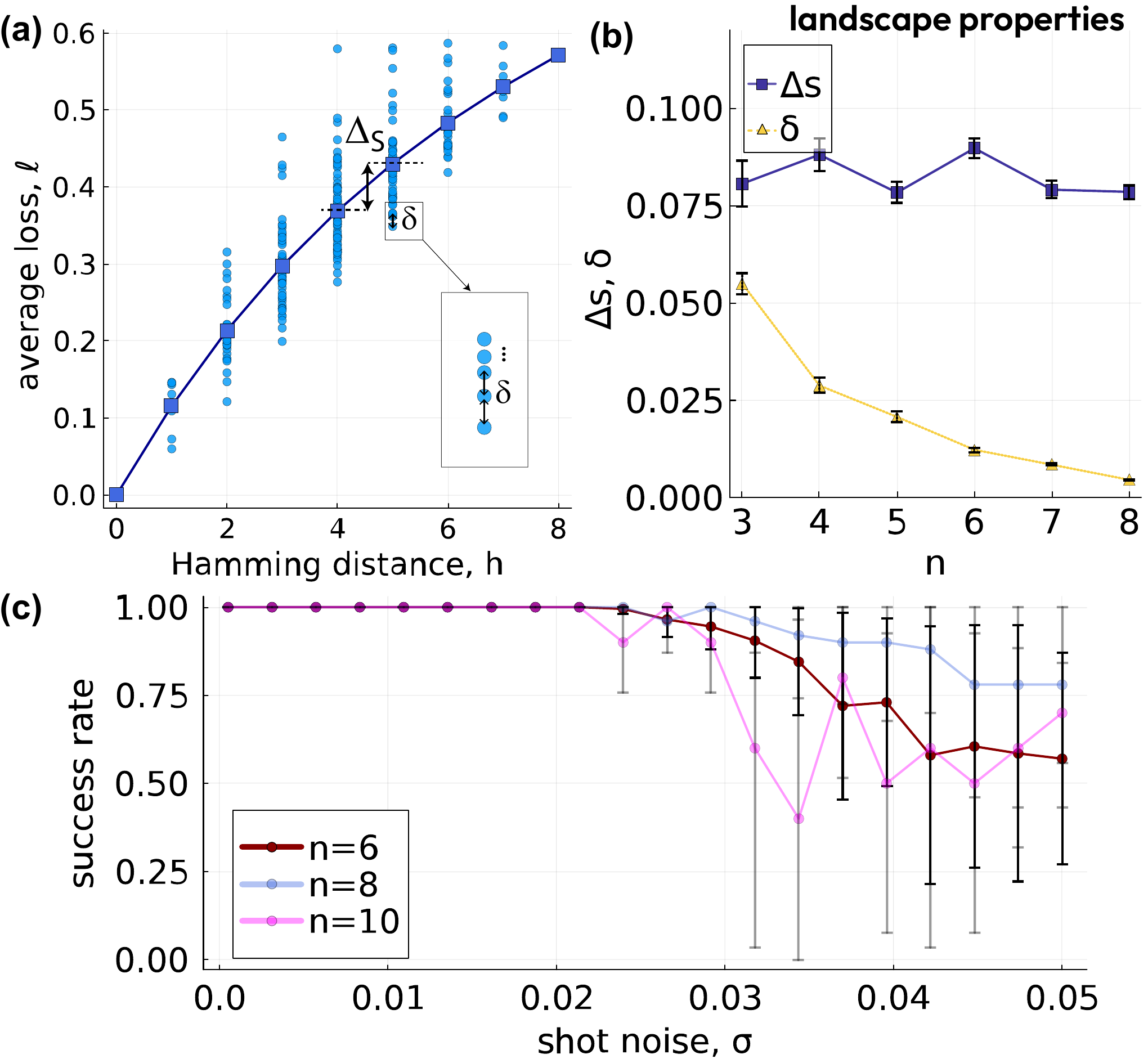}
    \caption{\textbf{Noisy discrete optimization.} \textbf{(a)} Loss vs Hamming distance to optimal solution, for $n=D=8$, $\beta_{\mathrm{W}} = \beta_{\mathrm{V}}=0.2$, shows a clear trend (the average value of $\ell_a$ is shown by the purple curve). Relevant properties of the landscape (distance between mean values $\Delta_{\mathrm{S}}$ and gaps $\delta$) are visualized. \textbf{(b)} Properties of the loss landscape that help to explain the success of optimization for landscapes with small gaps between individual loss values. The single local step loss difference $\Delta_S$, which determines the viability of hill climbing, remains constant with $n$. In contrast $\delta$, which does not matter for hill climbing but instead roughly quantifies the loss difference between random landscape points, shrinks with $n$. \textbf{(c)} Hill climbing under finite number of shots, showing an average success rate for the given standard deviation $\sigma$ (defined by number of shots per evaluation). Results are shown for problems with $n=D=6,8,10$, and $\beta_{\mathrm{W}} = \beta_{\mathrm{V}}=0.2$.}
    \label{fig:noisy}
\end{figure}

Here, we visualize the scaling of key parameters of a typical loss landscape of the devised recompilation problem. We show one example in Fig.~\ref{fig:noisy}(a), highlighting its key properties. These correspond to: 1) difference of mean loss between neighboring Hamming distances, $\ell(h) - \ell(h-1)$, averaged over all $h$ (denoted as sliding step $\Delta_{\mathrm{S}}$); and 2) average distance between ordered loss values for a given Hamming distance $h=\lceil D/2 \rceil$, which we call $\delta$. Of these, the key property for understanding exponential concentration is $\Delta_{\mathrm{S}}$, as this determines the loss differences observed between adjacent variables when running the hill climbing algorithm. In particular, we are interested in the scaling of $\Delta_{\mathrm{S}}$ as we increase $n$ and $D$. The parameter $\delta$ quantifies the typical change in loss for a random global change in parameters. In a sense, it quantifies the average flatness of the landscape but crucially does not affect the trainability via hill climbing. 

The results are shown in Fig.~\ref{fig:noisy}(b), averaged over various configurations for increasing system size $n=D$. Crucially, the sliding step $\Delta_{\mathrm{S}}$ shows that not only are loss differences not exponentially concentrated, but in fact, little changes with the problem size. 
We highlight that these results are for $\beta = 0.2$ and $D = n$: we will later see that this corresponds to a high entanglement and magic regime. 
The $\delta$ parameter does decrease with system size, indicating that for randomly chosen points the landscape becomes flatter, but this does not affect the trainability via local adaptive methods.

Having established that loss differences do not vanish exponentially with system size, we now proceed to test hill climbing-based circuit recompilation in the case of finite number of measurement shots, where the shot noise makes loss evaluation stochastic. 
In the presence of shot noise, the loss estimator $\ell$ fluctuates with a standard deviation $\sigma \approx 1/\sqrt{N_{\mathrm{shots}}^{(1)}}$, where $N_{\mathrm{shots}}^{(1)}$ is the number of measurement shots allocated per trial (single loss evaluation). The observed loss can be approximately modeled as $\tilde{\ell} = \ell + \sigma \, \eta$, where $\eta \sim \mathcal{N}(0,1)$ is a random variable sampled from the normal distribution $\mathcal{N}(0,1)$ multiplied by the standard deviation $\sigma$. 
To account for this noise, we modify the hill climbing procedure by adapting both the number of neighbors explored (the search breadth $\lambda$ \cite{doerr2018discrete}) and the number of trials used per neighbor (depth $m$), according to the current estimated loss $\tilde{\ell}$ (see the description in Appendix~\ref{sec:noisy_hill_climbing}). This adaptive schedule allows for efficient noisy hill climbing with the same number of total function evaluations as in the noiseless case, for a suitably chosen $\sigma$.  
%%%
%%%

The results for hill climbing with shot noise are shown in Fig.~\ref{fig:noisy}(c). We perform simulations for different system sizes and problems, initial bitstrings, and noise realizations, and quantify success rate $p_{\mathrm{succ}}$ as a fraction of correct solutions. The adaptive schedule remains the same. Here, stopping criteria are set based on noisy loss evaluations, while $p_{\mathrm{succ}}$ is evaluated by comparing $s_{\mathrm{final}}$ and $s^*$ after each trial. We observe that success rate of noisy hill climbing remains perfect until the shot noise level reaches $\sigma \approx 0.05$. Importantly, this is true for problems growing from $n=D=6$ to 8 and 10 qubits, without major changes in performance. 
This happens despite the significant growth of the solution space and the decrease in distances between loss values.
We conjecture that as long as the number of shots is chosen such that the standard deviation $\sigma$ is kept below half of the mean separation $\Delta_{\mathrm{S}}$, the optimization procedure remains robust to loss fluctuations, and the iterative procedure follows the (on average) unimodal landscape. \\

%---------------------------

\paragraph{Is the problem robust to surrogating strategies via classical shadows?} It has been highlighted that classical shadows can sometimes be used to first create a surrogate of a quantum landscape, before a variational procedure is applied to this surrogate model~\cite{basheer2022alternating, jerbi2023power, cerezo2023does, lerch2024efficient, bermejo2024qcnns}. In other words, a problem may be tackled via quantum-enhanced classical simulation methods, avoiding an adaptive hybrid quantum-classical loop. While we cannot entirely rule out exotic ``measure-first" schemes that do not resemble current known techniques, we now present evidence that our proposed problem structure and input state cannot be surrogated via standard approaches using Pauli and Clifford shadows~\cite{Huang2020shadows}.

We start by noting that Pauli shadows can only be used to solve an optimization problem if the optimization problem can be solved using local information. If the problem involves an input state which is a highly entangled `volume law' state, then the local reduced states of the system will be exponentially close to maximally mixed. In this case, exponentially many resources are required to extract any information that can be used to solve the target problem, and so Pauli shadows are not useful. 

\begin{figure}[t!]
\includegraphics[width=1.0\linewidth]{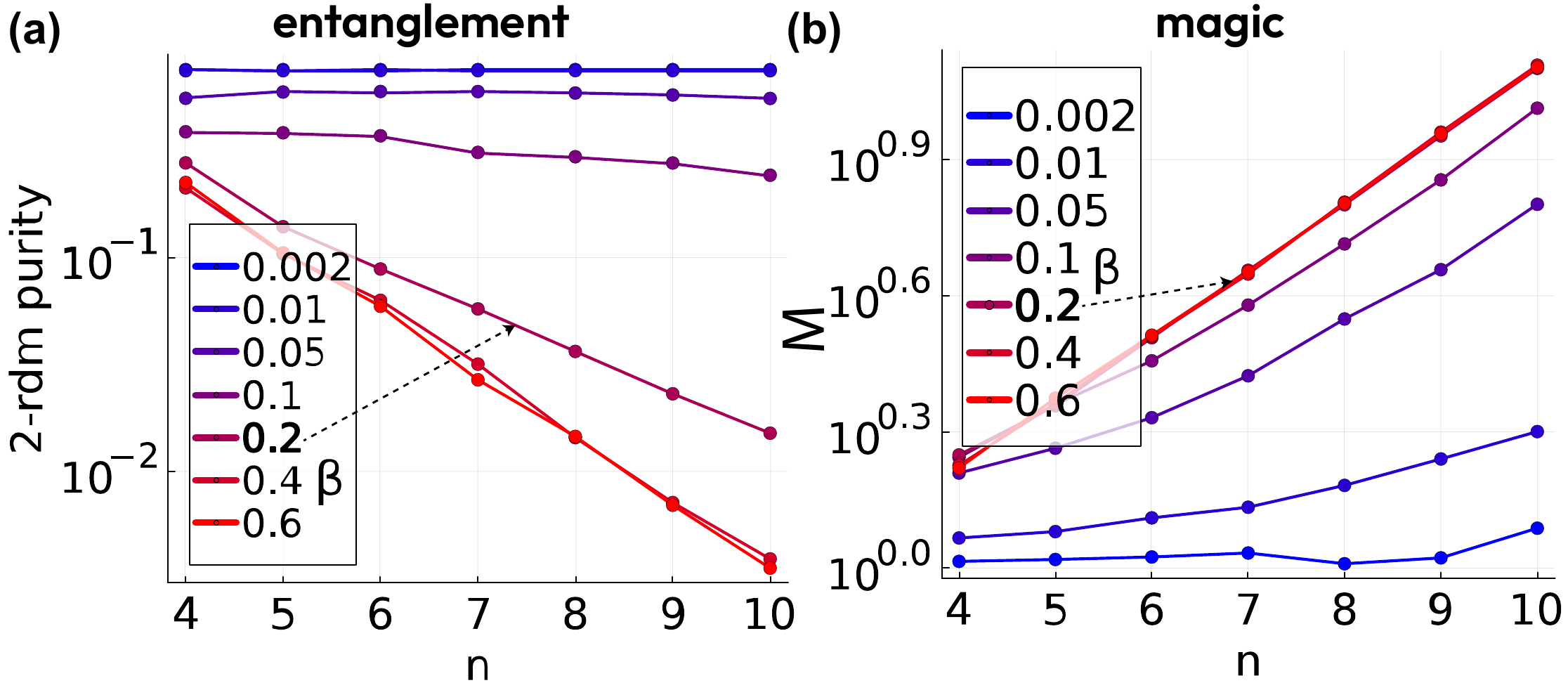}
    \caption{\textbf{Hardness of puzzle circuits.} \textbf{(a)} Entanglement properties of different puzzle circuits for $n = D$, showing the purity of 2-body reduced density matrices after subtracting 1/4. This is equivalent to the squared two-norm distance to the maximally mixed state. Scaling for different $\beta_{\mathrm{W,V}}$ (per individual block) is shown. \textbf{(b)} The Stablizer norm $M$ (a lower bound on the robustness of magic, a non-Cliffordness measure) is shown as a function of $n$ for different $\beta$. Our previous sweet spot of $\beta = 0.2$ is highlighted in bold.}
    \label{fig:hardness}
\end{figure}

Here we show that the input data states, for the same parameter regimes for which we previously found that the landscape differences do not exponentially concentrate with system size (namely, $\beta_{\mathrm{W}} = \beta_{\mathrm{V}} = 0.2$ and $D =n$), are highly entangled volume law states \cite{Hoke2023} (and so cannot be efficiently characterized using Pauli shadows). In particular, in Fig.~\ref{fig:hardness}(a) we plot the average distance of the reduced two-qubit states of the target to the maximally mixed state as a function of system size (we take the mean over different partitions and simulate five instances in each case). We find that this distance vanishes exponentially in $n$ for increasing $\beta$, with a steep exponential decline at our previous sweet spot of $\beta = 0.2$. In Appendix~\ref{sec:partially_random_properties} we provide additional tests that further demonstrate properties of partially-random unitaries and quantify the onset of volume-law entanglement~\cite{Nakagawa2018}.
%%%
%%%

Next, we note that Clifford shadows only work if we have an efficient classical representation of the states that we wish to compute the overlap with~\cite{Bertoni2024}. Typically, this will be in the form of a stabilizer state or, more generally, has a known approximation in terms of a polynomial in $n$ number of stabilizer states. This would be the case if the circuit to prepare the state uses only a logarithmic in $n$ number of T gates. Our ansatz states, however, contain $\Theta(n)$ T gates for the puzzles and many more in the construction of the scrambling unitaries. Thus it seems highly unlikely that there is an efficient classical representation of these states in terms of stabilizer states. Nonetheless, to further support these claims we show that the target states, and therefore also good guesses for this target state, are high magic states and therefore cannot be efficiently represented by stabilizer states. 

A natural magic measure to probe the efficiency of representing a state in terms of stabilizers is given by the \textit{robustness of magic} which is the minimal 1-norm of a quasi-probability decomposition of a state into pure stabilizer projectors~\cite{howard2017application}. However, due to the minimization, this measure is computationally inefficient to compute. Instead, as is common in the literature, we make use of the fact that the `stabilizer norm'~\cite{campbell2011catalysis} lower bounds the robustness~\cite{howard2017application}. This measure is defined as 
\begin{equation}
M(\psi) = \frac{1}{2^{n}} \sum_{P \in \mathcal{P}_{n}} \big|\langle \psi | P | \psi \rangle\big| ,
\label{eq:M1}
\end{equation}
where $\mathcal{P}_{n} = \{I, X, Y, Z\}^{\otimes n}$ can be averaged over the full set of Pauli strings for smaller $n$, or sampled (updating the normalization accordingly).

Results are presented in Fig.~\ref{fig:hardness}(b) for increasing $n$ and $\beta_{W,V}$ growing from $0.002$ to $0.6$, and averaged over five instances in each case. We observe that magic grows with $\beta$, and already at $\beta=0.2$ magic grows exponentially fast in $n$. Thus this is further evidence that Clifford shadows would seemingly not be computationally efficient for this problem. We extend our characterization of magic for partially-random circuits in Appendix~\ref{sec:partially_random_properties}, where typical stabilizer fidelity is probed \cite{Bravyi2019simulationofquantum}. \\

%-------------

\paragraph{Large scale numerical tests.} So far we have tested the variational recompilation for smaller systems, limited by the numerical simulation complexity when working with random operators. To test discrete optimization at larger scale \textit{in silico} we need to relax conditions leading to circuit simulation hardness, either by reducing magic or reducing entanglement. We choose the latter and showcase that trainable landscapes can be generated at a scale of hundreds of qubits. Note that with this, we cannot directly test the landscape used before, but rather we engineer a similar looking landscape where loss depends on correct placement of discrete blocks. For this, we introduce a family of circuits that instead of using randomness to reach high loss $\ell$, utilize layers of $R_Y$ rotations with randomized angles. Here, rotating the basis from $\{|0\rangle,|1\rangle\}$ to $\{|+\rangle,|-\rangle$) leads to a similar growth of the loss, and conjugated T gates contribute to this increase. Additional diagonal gates ($R_Z$ and sparse layers of $CZ$ gates) ensure that states are moderately non-trivial.

\begin{figure}[t]
\includegraphics[width=1.0\linewidth]{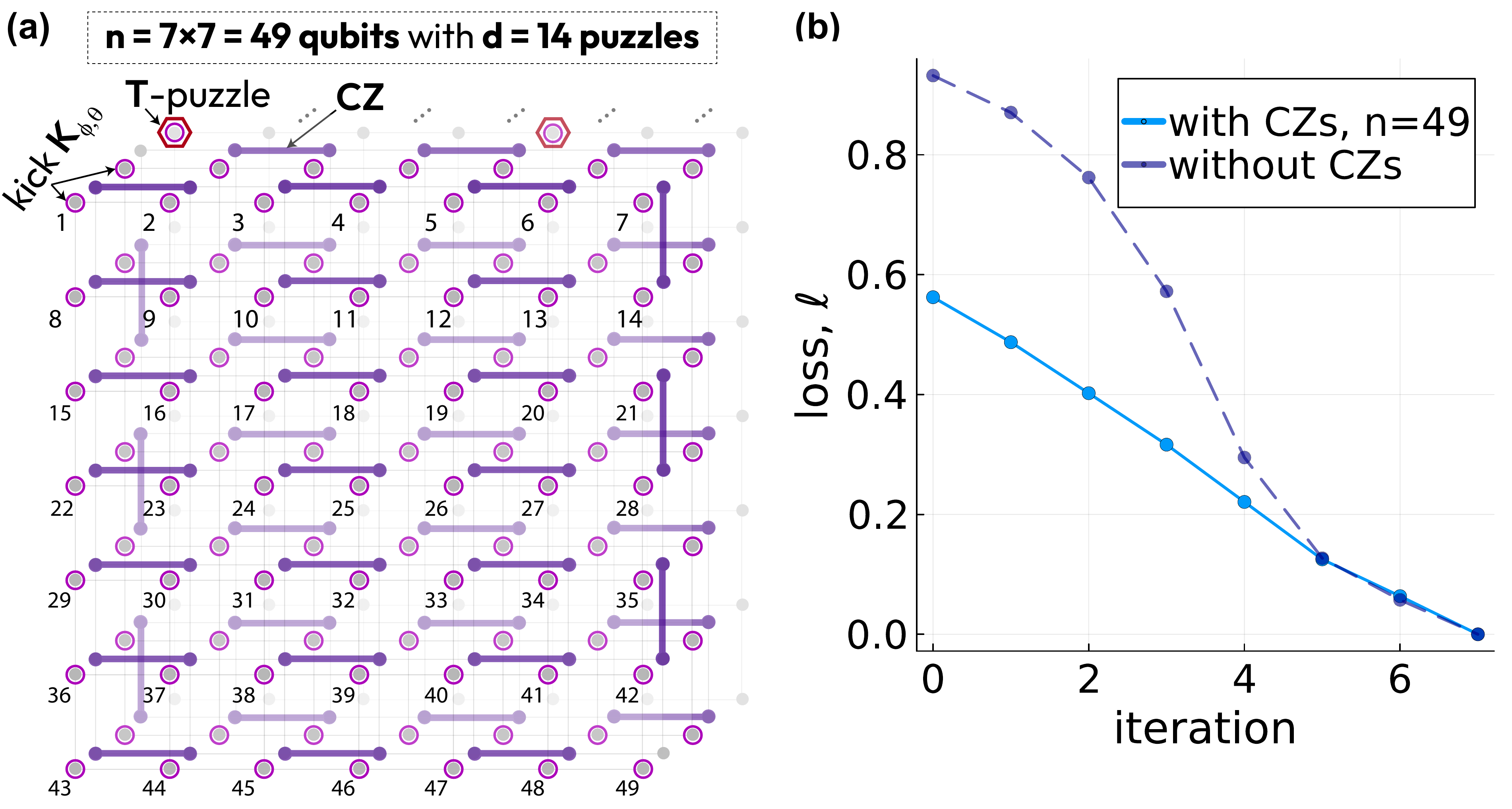}
    \caption{\textbf{Hill climbing at scale. (a)} Quantum circuit based on random single qubits rotations and sparse CZ layers (limited entanglement) that encode $D=14$ puzzles (locally-dressed T gates) on a $7 \times 7$ square grid. Circuit is contracted via 3D tensor network backend. \textbf{(b)} Discrete optimization run for $n=49$ with hill climbing applied for circuits with $CZ$ layers (blue curve). A simple instance without $CZ$s are shown for the reference by the dashed dark blue curve.}
    \label{fig:large-scale}
\end{figure}

We compile random instances of rotation-based partially-random circuits on a square lattice [Fig.~\ref{fig:large-scale}(a)]. A unitary $U(s^*)$ has the same form as in Eq.~\eqref{eq:Us}, where scrambling operators are constructed as $W_i = (\prod_{j,j' \in \mathcal{L}_2}CZ_{j,j'})\cdot K_{\varphi',\theta';\forall} \cdot (\prod_{j,j' \in \mathcal{L}_1}CZ_{j,j'})\cdot K_{\varphi,\theta;\forall}$. Here, $K_{\varphi,\theta;\forall} := \bigotimes_{j=1}^n R_{Z,j}(\varphi_j)\,R_{Y,j}(\theta_j)\,R_{Z,j}^\dagger(\varphi_j)$ represents a kick operator on all qubits where angles $\{\varphi_j\}$ of single-qubit $R_Z$ rotations are chosen uniformly at random between $0$ and $2\pi$, and angles $\{\theta_j\}$ for $R_Y$ rotations are drawn from the normal distribution $\mathcal{N}(0,1)$ and multiplied by a scaling factor $\sigma_{\mathrm{rot}}$. The entangling layers of $CZ$ gates are applied to subsets of edges $\mathcal{L}_{1,2}$, corresponding to odd-pairs $\{(1,2), (3,4), \ldots, (47,48)\}$ and even-pairs $\{(2,3), (4,5), \ldots, (48,49)\}$, respectively [Fig.~\ref{fig:large-scale}(a)]. The dressing operators are applied as $V_i = (K_{\varphi,\theta;t[i]})^{L_V}$ where the kick is applied to qubits $t \in \mathcal{T}$ as selected targets, $|\mathcal{T}| = D$, matching those where puzzles are located ($CZ$ layers are omitted to simplify calculations; rotations can be applied $L_V$ times). There is a limited build-up of entanglement due to alternating bonds, but not full volume-law growth. The T gates are added as before (we choose $s^*=11\ldots1$ for simplicity) and the ansatz $\bar{U}(s)$ follows the same recipe, together with the measured observable and associated loss [Eq.~\eqref{eq:loss_s}].
%%%
%%%

We proceed to evaluate the circuit over the $7\times 7$ grid with $n=49$ qubits and $D=14$ puzzles, where target indices are spread around the lattice (more details provided as a code in \cite{github4code}). The loss is evaluated via general tensor network (TN) contraction using \texttt{Yao.jl} \cite{Luo2020yao} with the \texttt{yao2einsum} backend and \texttt{Greedy} contraction schedule. Similar 3D TN contraction was used in Ref.~\cite{Pan2022} to simulate Sycamore circuits \cite{Arute2019}. Hill climbing begins from a random string, with representative examples being $D/2$ Hamming distance away from all zero and ones. We set $\sigma_{\mathrm{rot},1} = 0.25$ for $W_i$ layers and $\sigma_{\mathrm{rot},2} = 0.4$ with $L_V=2$ for $V_i$ operations, approximately matching the $\beta$ parametrization of partially-random unitaries. 

The results are shown in Fig.~\ref{fig:large-scale}(b); we observe perfect recompilation. In the case with $CZ$ layers a total of $672$ two-qubit gates are applied and around $9000$ single-qubit gates. We note that the effective landscape remains sharp in the presence of entanglement (significant drop of $\ell$ at each iteration). This property shall also be instrumental in the presence of noise, albeit less relevant in the case of fault-tolerant circuit recompilation.

%-----------

\section{Discussion}

We have introduced a circuit recompilation task, where the goal is to identify a hidden configuration of puzzle gates embedded within partially-random circuits. The approach combines a classically hard-to-simulate input state with a trainable, discrete optimization landscape, making it well-suited for adaptive variational strategies. By searching over candidate bitstrings using a simple hill climbing algorithm, we demonstrated high recovery rates and robust convergence across a wide range of hyperparameters, valid for large system sizes. 

Our algorithm is based on a simple yet powerful observation: entanglement is not necessarily a roadblock for variational search, as its degree (also related to scrambling) can lead to a trainable loss landscape. Each block that we undo correctly has a meaningful contribution. However, due to the importance of the order in which we find the T gates and the potential for the T gates to interference, the landscape is non-separable and they cannot be independently optimized. Our results thus show that adaptivity (aka a hybrid quantum-classical feedback loop) is essential to the success of the protocol. In contrast, non-adaptive approaches, including measurement-first strategies, face exponential overheads and fail to resolve the hidden structure at scale. The problem therefore highlights a natural regime in which quantum adaptivity is not only useful, but necessary.

It is interesting to reflect on our findings in the context of the claims in Ref.~\cite{cerezo2023does} that variational quantum algorithms that provably avoid exponential concentration can be classically surrogated. 
As noted in the introduction, there are many differences between our settings, but potentially the most conceptually significant is that we consider a discrete optimization problem, whereas their case-by-case argument focuses on continuous variable problems. In particular, it is intriguing that the only other known counterexample is also a discrete optimization problem~\cite{cerezo2023does} (albeit one that relies on the standard `sew Shor into a QML' trick and is detached from any learning/optimization problem). It is thus natural to ask whether discrete optimization might more generally provide a setting to identify trainable variational quantum algorithms that cannot be classically surrogated.
This is particularly noteworthy given that, while currently less popular in quantum machine learning (QML), discrete optimization provides a natural setting for extending QML to fault-tolerant applications; avoiding parametrized rotations and instead working with blocks of operations \cite{chivilikhin2020mogvqe}. 

While the T gate recompilation problem has largely been constructed as a proof-of-principle in order to establish a separation between adaptive and non-adaptive access to quantum hardware, it is conceivable that it could be modified to more realistic use-cases. 
For example, the proposed approach is potentially relevant for early fault-tolerant architectures, where non-Clifford resources such as T gates are supplied via magic state distillation and teleportation \cite{Bravyi2005,nielsen2010quantum,SalesRodriguez2025,ZhouChang2000,DANOS2007,Morimae2012,Sano2022,Polacchi2023,Simmons2024,Main2025,baranes2025ftqcblind}. Specifically, it can be applied in the case where partial information about gate placement may be hidden from the compiler \cite{Akahoshi2024,Toshio2025,Bocharov2015} (see Appendix~\ref{sec:applications}). Similar optimization principles apply to Trotterized or QAOA-type dynamics \cite{Zhuk2024,ZhuEconomou2022,Chandarana2022}, where correct sequencing of unitaries must be discovered. We note that the value of adaptive recompilation is expected to be especially high for larger systems and fixed modular blocks (e.g. lattice surgery), and will further increase as we progress toward the full fault-tolerant regime.

%========================

\begin{acknowledgments}
O.K. thanks Elham Kashefi for interesting discussions on the subject. Z.H. thanks Truman Yu Ng and Maite Arcose for helpful discussions on measures of magic. O.K. and C.U. acknowledge the funding from UK EPSRC under award number EP/Z53318X/1 (QCi3 Hub) and EP/Y005090/1. Z.H. acknowledges support from the Sandoz Family Foundation-Monique de Meuron program for Academic Promotion.
\end{acknowledgments}

%\bibliography{bibliography, quantum}

\input{main.bbl}

\clearpage

\newpage

\appendix

\setcounter{figure}{0}
\renewcommand{\thefigure}{S\arabic{figure}}
\renewcommand{\theHfigure}{S\arabic{figure}}

\section{Implementing partially-random unitaries via the Cayley transform}\label{sec:Cayley_transform}

\subsection{Properties of the Cayley transform circuits}
In practice, partially-random unitaries can be implemented in various forms. One method relies on using the Cayley transform \cite{Cayley1846,helfrich2018orthogonalrnns,abanin2025chaos},
\begin{equation}
W(\beta) = \frac{\left( \mathds{1} - i \beta \mathds{H} \right)}{\left( \mathds{1} + i \beta \mathds{H} \right)},
\label{eq:cayley_transform}
\end{equation}
where the product of numerator and denominator involving the skew-Hermitian operator $i \mathds{H}$ ensures the unitarity of $W$. This operator is trivial at $\beta$ equal to zero, $W(0) = \mathds{1}$, while at increased $\beta > 1$ the unitary $W(\beta)$ becomes quasi-random, finally reaching another trivial point at $W(\beta \rightarrow \infty) = -\mathds{1}$. The operator $\mathds{H}$ is constructed as a random linear combination of Pauli strings, $\mathds{H} = (1/\sqrt{k}) \sum_{j=1}^{k} P_j$, where each $P_j$ is an off-diagonal tensor product of single-qubit operators $\{I, X, Y\}$. In principle, the number of strings $k$ can be as large as $3^n$, but in practice we can restrict it to polynomially many terms, $k = O(\mathrm{poly}(n))$, generating sufficient randomness (as verified in examples below). Since the Cayley transformation corresponds to the rational function $f(x) = (1 - i \beta x)/(1 + i \beta x)$, $x \in \mathbb{R}^+$, we show that this can be implemented on operator $\mathds{H}$ using a low-depth quantum singular value transformation (QSVT). We also propose a simple and efficient block-encoding for $\mathds{H}$, compiled at reduced cost in a nested form using single-qubit controlled strings $\{P_j\}$ (see discussion below). We note that due to the properties of $\mathds{H}$ and $f(x)$, the QSVT sequence implementing $f(x)$ applies the Cayley transformation \emph{deterministically} and with low-degree polynomials.

\subsection{Compiling the partially-random circuits}

To implement the Cayley transform $W(\beta)$ (Eq.~\eqref{eq:cayley_transform}), we first need to block encode the random Hermitian operator $\mathds{H} = (1/\sqrt{k}) \sum_{j=1}^{k} P_j$. To reduce the resource requirements, we take inspiration from quantum phase estimation (QPE) and its concatenation of operators at different powers \cite{kitaev1995qpe}. 

Typical QPE includes a controlled sequence made up of a cascade of unitaries $U^{2^m}$ acting on a system register, controlled by a single qubit $m$ from the ancilla register. With $K$ controlled operators, we can implement the operation $U_C = \sum_{x=0}^{2^K-1}|x\rangle\langle x| \otimes U^x$. In this way, the unitaries $\{U^{2^m}, m\in\{0,1,\ldots,K -1\}\}$ form a basis of size $K$ for the set of $k=2^K$ unitaries $\{U^x, x\in\{0,1,\ldots,k-1\}\}$.
%%%
\begin{figure}[t!]
    \centering
    \begin{quantikz}[wire types={q,q,n,q,b},classical gap=0.1cm, column sep = 0.3cm, row sep=0.2cm]
    \lstick[4]{$|0\rangle^{\otimes K}$}&\gate{H}&\ctrl{4}\gategroup[5,steps=4,style={rounded corners,fill=green!10, inner
    sep=2pt},background]{nested Paulis}&&&&\gate{H}&\meter{}\\
    &\gate{H}&&\ctrl{3}&&&\gate{H}&\meter{}\\
    \vdots &&&&\cdots&\vdots\\
    &\gate{H}&&&&\ctrl{1}&\gate{H}&\meter{}\\
    \lstick{$|\psi\rangle$}&&\gate[style={fill=red!20}]{B_0} & \gate[style={fill=red!20}]{B_1} &\ \cdots \ & \gate[style={fill=red!20}]{B_{K-1}}&& \rstick{$\mathds{H}|\psi\rangle$} 
    \end{quantikz}
    \caption{\textbf{Circuit $\hat{U}_{\mathds{H}}$ implementing the random Hermitian $\mathds{H}$}. Each basis element $B_i$ is applied on the system register, controlled by the $i$'th qubit of the ancilla register. A total of $K$ basis elements are needed to create a sum over $k=2^K$ Pauli operators; representing an exponential depth suppression in comparison to a standard LCU implementation. When $\mathbf{0}$ is measured in the ancilla register, $\mathds{H}$ is applied onto the system register (up to normalisation).}
    \label{fig:circuit_rand_hermitian}
\end{figure}
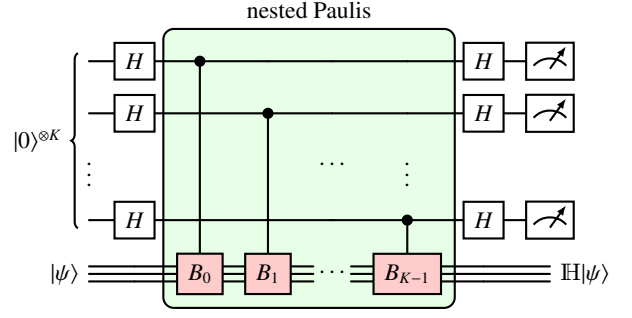

Based on this, we select a basis of mutually commuting Pauli strings $B=\{B_j\},j\in\{0,1,\ldots,K-1\}$ and $B_j\in\{\mathds{1},X,Y\}^{\otimes n} \ \forall j$. These Pauli operators form a basis of size $K$ for a set of $k=2^K$ Paulis $\{P_j, x\in\{0,1,\ldots,k-1\}\}$, with each Pauli $P_j$ formed from the product of a subset of the basis. We enforce the condition of commutativity as part of the method to ensure that each of these products of basis elements produces another Pauli with a coefficient of 1 (i.e. Hermitian). By interleaving this control sequence with layers of Hadamards on the ancilla register, we are able to prepare the random Hermitian operator as a normalized sum of Paulis (Fig.~\ref{fig:circuit_rand_hermitian}). In this way, the circuit $U_{\mathds{H}}$ can be referred to as a $K$-qubit block encoding of $\mathds{H}$ \cite{gilyen2019qsvt}: $\mathds{H}=\left(\bra{0}^{\otimes K}\otimes \mathds{1}\right)U_{\mathds{H}}\left(\ket{0}^{\otimes K}\otimes \mathds{1}\right)$ (up to normalization). To be specific, $U_{\mathds{H}}$ encodes the operator $(1/k) \sum_{j=1}^{k} P_j$, but we can absorb the extra $\sqrt{k}$ normalisation factor into the Cayley transform itself; $\beta \rightarrow\beta\sqrt{k}$. 
%%%
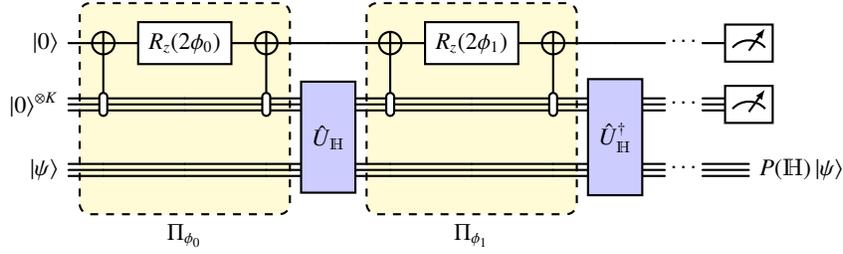
\begin{figure*}[t!]
   \centering
   \begin{quantikz}[wire types={q,b,b}, classical gap=0.08cm, row sep=0.2cm, column sep=0.3cm]
   \lstick{$\ket{0}$} & \targ{}\gategroup[3,steps=3,style={dashed,rounded corners,fill=yellow!20, inner xsep=1.5pt},background,label style={label position=below,anchor=north,yshift=-0.2cm}]{{$\Pi_{\phi_0}$}} & \gate{R_z(2\phi_0)} & \targ{} && \targ{}\gategroup[3,steps=3,style={dashed,rounded corners,fill=yellow!20, inner xsep=1.5pt},background,label style={label position=below,anchor=north,yshift=-0.2cm}]{{$\Pi_{\phi_1}$}} & \gate{R_z(2\phi_1)} & \targ{} &&\ \cdots \ & \meter{} \\
   \lstick{$\ket{0}^{\otimes K}$} & \octrl{-1} && \octrl{-1} & \gate[wires=2,style={fill=blue!20}]{\hat{U}_{\mathds{H}}} &\octrl{-1} && \octrl{-1} & \gate[wires=2,style={fill=blue!20}]{\hat{U}_{\mathds{H}}^\dagger}& \ \cdots \ & \meter{}\\
   \lstick{$\ket{\psi}$} &  &&  &&  &&  &&\ \cdots \ & \rstick{$P(\mathds{H})\ket{\psi}$}
   \end{quantikz}
   \caption{\textbf{Snippet of the QSVT sequence} $U(\boldsymbol{\phi},\mathds{H})$ for the transformation $\mathrm{H}\rightarrow P(\mathds{H})$, with the circuit for $\hat{U}_{\mathds{H}}$ given in Fig.~\ref{fig:circuit_rand_hermitian}. Since $P(\mathds{H})$ is unitary, $\mathbf{0}$ is measured with 100\% probability on the ancilla qubits. This results in the transformed matrix $P(\mathds{H})=W(\beta)$ being applied deterministically onto the system register $\ket{\psi}$. The accuracy of the application of the Cayley function depends on the choice of angles $\boldsymbol{\phi}$.}
   \label{fig:QSVT_implementation}
\end{figure*}
%%%

To perform the Cayley transform on $\mathds{H}$, we make use of the quantum singular value transformation (QSVT) procedure \cite{gilyen2019qsvt, martyn2021unification}, which allows for the implementation of a polynomial transformation on the singular values of the target matrix. QSVT is a generalization of quantum signal processing (QSP), which applies polynomial transformations to scalars. QSP relies on the interleaving of two types of single-qubit rotation, 
\begin{equation}\label{eq:QSP}
    \begin{aligned}
        U(\boldsymbol{\phi},x)&=e^{i\phi_0 Z} \prod_{k=1}^{d} \left( W(x) e^{i\phi_k Z} W(x)^\dagger \right).\\
        &=\begin{bmatrix}
        P(x) & iQ(x)\sqrt{1-x^2}\\
        iQ^*(x)\sqrt{1-x^2} & P^*(x) 
    \end{bmatrix},
    \end{aligned}
\end{equation}
where $W(x)$ is a signal rotation operator that encodes the scalar $x$, and $\boldsymbol{\phi}$ is a vector of phase angles $\boldsymbol{\phi}=\{\phi_0,\phi_1,\ldots \phi_d\}$. This sequence with the correct set of angles performs the transformation $x \mapsto P(x)$, where $P(x) = \langle0|U(\boldsymbol{\phi},x)|0\rangle$. The complex polynomials $P(x)$ and $Q(x)$ have several constraints: 1) $\deg(P) \leq d$, $\deg(D) \leq d-1$; 2) $P$ has parity $d\sMod 2$ and $Q$ has parity $(d-1)\sMod 2$; and 3) $|P|^2+(1-x^2)|Q|^2=1$.

These constraints are crucial in the analysis of our function of interest, $f(x) = (1 - i \beta x)/(1 + i \beta x)$. We can see that this is a complex function, with $\mathrm{Re}[f(x)]$ being an even function and $\mathrm{Im}[f(x)]$ an odd function. Taking $\mathrm{Re}[P(x)]=\mathrm{Re}[f(x)]$ representing our target function, constraint 3) fixes $|P(\pm1)|=1$. Constraint 2) tells us that if $P(x)$ is an even function, then $Q(x)$ must be odd. Therefore, $Q(0)=0$, and hence constraint 3) also enforces that $|P(0)|=1$.

It is also important to note that $|f(x)|=1 \ \forall x$. Therefore, if we find an accurate sequence of angles $\boldsymbol{\phi}$ such that $\mathrm{Re}[P(x)]=\mathrm{Re}[f(x)]$, the normalization conditions at $x\in\{-1,0,1\}$ will automatically enforce $\mathrm{Im}[P(x)]=\mathrm{Im}[f(x)]$. Hence the QSP function $P(x)$ represents the full complex function $f(x)$ for $x\in\{-1,0,1\}$. 
The target function can be expanded using a Taylor series; $\mathrm{Re}[f(x)]=(1-\beta^2x^2)/(1+\beta^2x^2)=1+\sum_{m=1}^\infty(-1)^m2\beta^{2m}x^{2m}$. This decomposition illustrates why a higher QSP degree $d$ is needed to represent $f(x)$ accurately when $\beta$ is large. The dependence of $d$ on $\beta$ is weak; from numerical tests, $D=4$ is enough to approximate $f(x)$ within an error $\epsilon\sim10^{-6}$ for $\beta\lesssim 0.75$, and even at $\beta=2$, we find that $D=10$ is sufficient.

Now, let us consider the transformation of $\mathds{H}$, which can be decomposed using the singular value decomposition (SVD) as $\mathds{H} = \hat{W}\Sigma\hat{V}$.  Here, $\hat{W}$ and $\hat{V}$ are unitary matrices and $\Sigma$ is a diagonal matrix containing the singular values of $\mathds{H}$, $\sigma_i = \Sigma_{ii}$. For an even function, we can use QSVT to perform the transform $\mathds{H}\mapsto P(\mathds{H})$ with 
\begin{equation}
    P(\mathds{H}) = \sum_{i=0}^{2^n-1} P(\sigma_i)\ket{v_i}\bra{v_i},
\end{equation}
where $\{\ket{w_i}\}$ and $\{\ket{v_i}\}$ denote the columns of $\hat{W}$ and $\hat{V}$ respectively. 
From this we can see that $P^\dagger(\mathds{H})P(\mathds{H}) = \sum_{i=0}^{2^n-1} P^*(\sigma_i)P(\sigma_i)\ket{v_i}\bra{v_i}$. The key observation is that for our operator $\mathds{H}$, $\sigma_i\in\{0,1\}\ \forall i$. This means that $P^*(\sigma_i)P(\sigma_i)=|P(\sigma_i|^2=1 \ \forall i$ and $P^\dagger(\mathds{H})P(\mathds{H}) = \sum_{i=0}^{2^n-1}\ket{v_i}\bra{v_i}=\mathds{1}$. Hence, for this type of operator, $P(\mathds{H})$ is a unitary transformation, and $P(\sigma_i)=f(\sigma_i)$ for $\sigma_i \in {0,1}$. Therefore $P(H)=W(\beta)$ is the exact unitary Cayley transform that we require.

As for the implementation, the QSVT sequence to implement $P(H)$ consists of layers of the block encoding $U_{\mathds{H}}$ alternating with projector-controlled phase (PCP) gates $\prod_\phi$. The PCP gate has the effect of applying $e^{i\phi}$ onto the subspace containing $\mathds{H}$  (the $n$ system qubits), while applying $e^{-i\phi}$ on the $K$-qubit ancilla subspace. This gate can be implemented using an additional ancilla qubit. We depict a snapshot of the QSVT circuit in Fig.~\ref{fig:QSVT_implementation}.

The QSVT sequence $U(\boldsymbol{\phi},\mathds{H})$ block encodes the polynomial transformation of the matrix: $P(\mathds{H})=\left(\bra{0}^{\otimes K+1}\otimes \mathds{1}\right)U(\boldsymbol{\phi},\mathds{H})\left(\ket{0}^{\otimes K+1}\otimes \mathds{1}\right)$. By applying $U(\boldsymbol{\phi},\mathds{H})$, then measuring $\mathbf{0}$ on the ancilla register, the transformed matrix $P(\mathds{H})$ is applied on the system register. 

Since $P(H)$ is a unitary, the success probability of projecting onto $\mathbf{0}$ on the ancilla register is $||P(\mathds{H})|\psi\rangle||^2 = 1$. Therefore, QSVT applied on the block encoding of $\mathds{H}$ gives us a deterministic protocol for applying the Cayley transform $W(\beta)$.

%---------------------------

\subsection{Properties of partially-random circuits}\label{sec:partially_random_properties}

We proceed to provide some intuition on why the described problem can be optimized successfully, while not being amenable to measure-first attacks. Here, let us consider a problem instance and corresponding circuit $R := \bar{U}(s)U(s^*)$ where $s \neq s^*$ and the Hamming distance $h(s,s^*) \gg 1$, defined as a minimal number of bit flips from $s$ to $s^*$. In this case $R$ represents a product of partially-random circuits, leading to a correspondingly high loss. In essence, we can see this as one partially-random circuit with large $\beta_{\mathrm{eff}} := \beta_{\mathrm{W}} d + \xi(\beta_V, d)$, which corresponds to a \emph{total} circuit depth, including partially-random operators $\{W_i\}_{i=1}^d$ as well as extra depth $\xi(\beta_V, d)$ arising from uncontracted dressed operators $\{V_i^\dagger \mathrm{T} V_i\}_{i=1}^d$. In the following we study different properties of such blocks, omitting the full puzzle structure for brevity.

\begin{figure}[t!]
\includegraphics[width=1.0\linewidth]{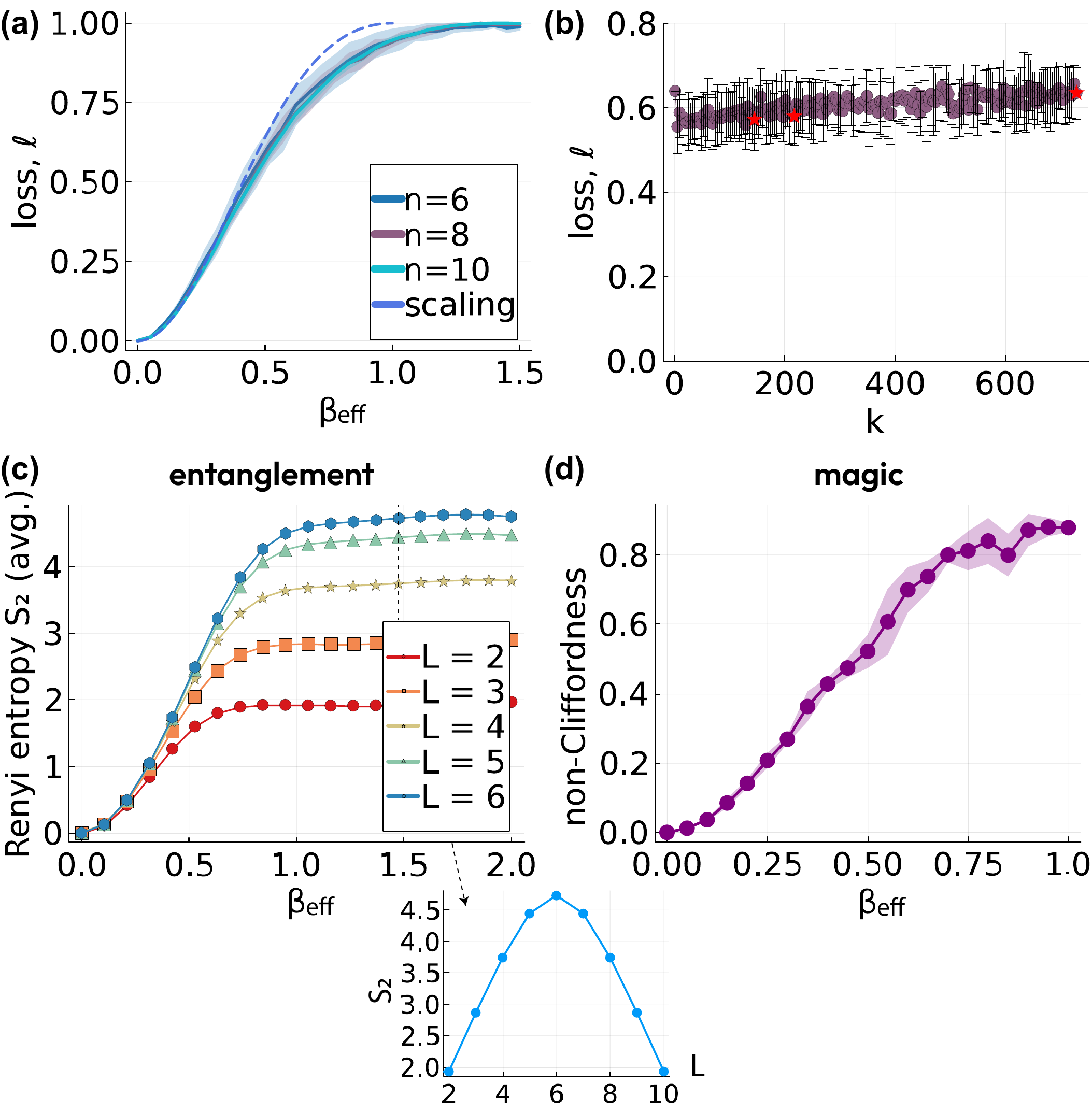}
    \caption{\textbf{Properties of partially-random circuits. (a)} Increase of the zero projector expectation value as a function of randomness parameter $\beta_{\mathrm{eff}}$, shown over 50 instances. Results for $n=6,8,10$ show that the landscape remains similar for increasing $n$. \textbf{(b)} Loss $\ell$ for different size of random Hermitian operators, where $k$ is a number of uniformly drawn Pauli strings for $n=6$. Red stars point to $k=4n^2$, $n^3$, and $3^n$. \textbf{(c)} Growth of 2nd-order Renyi entropy $S_2$ with increasing $\beta_{\mathrm{eff}}$ (applied as one big block), averaged over partitions of different size from $L=2$ to $L=6$ out of $n=12$ ($k=4n^2$), and demonstrating the onset of volume-law entanglement. The inset shows the scaling of entanglement entropy with $L$ at $\beta_{\mathrm{eff}} = 1.5$. \textbf{(d)} Non-Cliffordness parameter (proxy to magic) measured as a minimal infidelity of $|\psi_{\beta_{\mathrm{eff}}}\rangle$ and a set of states prepared by Clifford circuits sampled for $n=6$ (again, for a full circuit against $\beta_{\mathrm{eff}}$).}
    \label{fig:properties}
\end{figure}

In Fig.~\ref{fig:properties}(a) we show that zero projector's expectation $\ell(\beta_{\mathrm{eff}}) = 1 - |\langle \mathbf{0} |W(\beta_{\mathrm{eff}})|\mathbf{0} \rangle|^2$ grows quadratically at small fractions $\beta_{\mathrm{eff}}$, and saturates to its maximal value of one past $\beta_{\mathrm{eff}} = 1.0$. Statistics are collected over $50$ realizations for $30$ values of $\beta$, at different $n$. Importantly, this behavior is shared for increasing system size $n$, and helps to interpret results --- each correctly guessed bit leads to circuit contraction and (on average) lowers the loss at any system size for $0 < \beta_{\mathrm{eff}} < 1$. (Note that here we refer to the expected value $1-|\mathbf{0}\rangle \langle \mathbf{0}|$ as loss, even though no optimization is performed in this case.) The scaling can be fitted as $\ell(\beta_{\mathrm{eff}}) \approx 1 - ((1-\beta_{\mathrm{eff}}^2)/(1 + \beta_{\mathrm{eff}}^2))^2$, as motivated by the scaling of the Cayley transform and exponentially vanishing overlap of $W(\beta_{\mathrm{eff}})|\mathbf{0} \rangle$ and $|\mathbf{0} \rangle$.

Second, we test the dependence of the loss on the number of Pauli strings $k$ in $\mathds{H}$. The corresponding value is shown in Fig.~\ref{fig:properties}(b), where we observe a quick saturation of $\ell(\beta_{\mathrm{eff}},k) \approx 0.6$ at $\beta_{\mathrm{eff}} = 0.5$ past $k = O(n^2)$. Results are shown for $n=6$, with the three red stars in Fig.~\ref{fig:properties}(b) highlighting $k = 4n^2$, $n^3$, and $3^n$. This confirms that $\mathds{H}$ with $O(n^2)$ terms used within the Cayley transform can induce sufficient changes in loss for each block (as used in the main text throughout).
%%%
%%%

Third, we probe entanglement properties of the corresponding quantum data $|\psi_{\mathrm{in}}\rangle$ based on partially-random circuits with $\beta_{\mathrm{eff}} > 0.5$. For this, we characterize entanglement properties of states $|\psi_{\beta_{\mathrm{eff}}}\rangle := W(\beta_{\mathrm{eff}})|\mathbf{0} \rangle$ at increasing $\beta_{\mathrm{eff}}$ for $n = 12$ and different partitions. Specifically, we estimate the average Renyi entropy of order two, $S_2(\beta_{\mathrm{eff}};L) = -\log_2(\mathrm{tr}\{\rho_{\mathcal{A}}^2\})$, with results averaged over reduced density operators with different partitions $\{\mathcal{A},\mathcal{B}\}$ of length $L$ and $n-L$, respectively. In Fig.~\ref{fig:properties}(c) we observe that Renyi entropy increases with $\beta_{\mathrm{eff}}$. Importantly, for larger fractions of random operations ($\beta_{\mathrm{eff}} > 0.5$) we observe the linear increase of $S_2(\beta_{\mathrm{eff}} \approx 1; L)$ as a function of $L$, signifying the onset of volume law entanglement \cite{Nakagawa2018,Hoke2023}. We also plot the inset for Fig.~\ref{fig:properties}(c) showing the typical dome-like scaling of entanglement entropy with $L$ at $\beta_{\mathrm{eff}} = 1.5$. This shows that variational approaches can indeed work with highly entangled data.

Finally, we test the magic of the used quantum data by estimating its minimal infidelity with respect to stabilizer states generated from Clifford circuits. For a given partially-random state $|\psi_{\beta_{\mathrm{eff}}}\rangle$, we generate a set of stabilizer states $\{|\phi_j\rangle\}$ by applying randomly sampled Clifford unitaries to $|\mathbf{0}\rangle$. The stabilizer fidelity is defined as $F_{\mathrm{stab}}(\psi) = \max_j |\langle \phi_j | \psi \rangle|^2$ \cite{Bravyi2019simulationofquantum}. The non-Cliffordness measure is then $N(\beta_{\mathrm{eff}}) = 1 - F_{\mathrm{stab}}(\psi_{\beta_{\mathrm{eff}}})$, which vanishes for stabilizer states and increases as $|\psi_{\beta_{\mathrm{eff}}}\rangle$ departs from the stabilizer manifold. This provides an operational quantifier of the deviation from the Clifford-simulable subspace and helps assess the hardness of classical simulation. In Fig.~\ref{fig:properties}(d) we show magic sampled over $1000$ depth-$5$ Clifford circuits for $n=6$, and averaged over $5$ realizations. This further demonstrates state complexity past $\beta_{\mathrm{eff}} = 0.5$.

%-------------

\section{Hill climbing with noisy loss evaluation}\label{sec:noisy_hill_climbing}

In the presence of shot noise, evaluating a single neighbor during the hill climbing procedure is typically not sufficient to reliably identify an improving direction, since the estimated loss values $\tilde{\ell}$ fluctuate around their true means. However, for favorable landscapes where there is a clear (yet non-monotonic) direction, one can gather more information from neighboring configurations. To this end, one can introduce a breadth parameter $\lambda$, which sets the number of randomly chosen neighboring bitstrings that are tested at each iteration \cite{DoerrNeumann2020}. These $\lambda$ neighbors form a small \emph{population} of candidates, each evaluated with a limited number of trials $m$, each measured with $N_{\mathrm{shots}}^{(1)}$. The purpose is not to resolve every candidate accurately, but to capture the trends of the loss function. The decision rule is then based on the relative ordering of these population estimates. If the lowest observed neighbor loss $\tilde{\ell}_{\min}$ is smaller than the re-estimated current loss $\tilde{\ell}_{\mathrm{cur}}$ by a margin larger than the expected statistical fluctuation, the bitstring is updated to this best neighbor. Otherwise, the current configuration is retained and the search continues.

When $\tilde{\ell}$ is large, a broad search with small $m$ is preferred, which allows the algorithm to identify a descending trend despite the fact that individual evaluations are noisy, and following the global trend of the unimodal landscape with minimal trials per candidate. As $\tilde{\ell}$ becomes small (close to convergence), the breadth is reduced while the number of trials per evaluation is increased, so that candidate improvements are distinguished reliably near convergence. At each step, the current configuration is re-evaluated to avoid systematic bias, and a move is accepted only if the loss decrease is statistically significant. The stopping condition is when the optimum (zero loss) is reached with the noise level, or the number of allowed iterations is maxed out (set to $200$ in simulations).

We note that the noisy hill climbing algorithm converges to the true optimum with high probability, and generally requires a similar number of iterations but a slightly higher number of shots compared to the noiseless case. The adaptive allocation of breadth and depth helps to balance exploration and precision across the landscape. This is given by a function $\{\lambda, m\}(\ell)$, which produces suitable hyperparameters based upon the current best loss value $\ell$. The total number of trials is $\lambda \, m$. Specifically, we use the following piecewise schedule: $\ell \geq 0.3:~ (\lambda,m) = (36,1)$; $0.1 \leq \ell < 0.3:~ (\lambda,m) = (12,3)$; $\ell < 0.1:~ (\lambda,m) = (2,6)$. For $n=D=8$ shown in the main text, this shows that efficient updates are possible by studying even a small part of the total population. For instance, given random starting configurations and setting $\sigma = 0.02$ we converge with unity probability across varying system and problem sizes (Fig.~\ref{fig:noisy}(c) of the main text). We note that this holds despite diminishing gaps and growing number of configurations, concluding that the overall shape of the landscape (characterized by step size $\Delta_S$) is more important than the fine details within each step (granularity $\delta$).

%-------------

\section{Possible applications of circuit learning within FTQC}\label{sec:applications}

\begin{figure}[b!]
\includegraphics[width=1.0\linewidth]{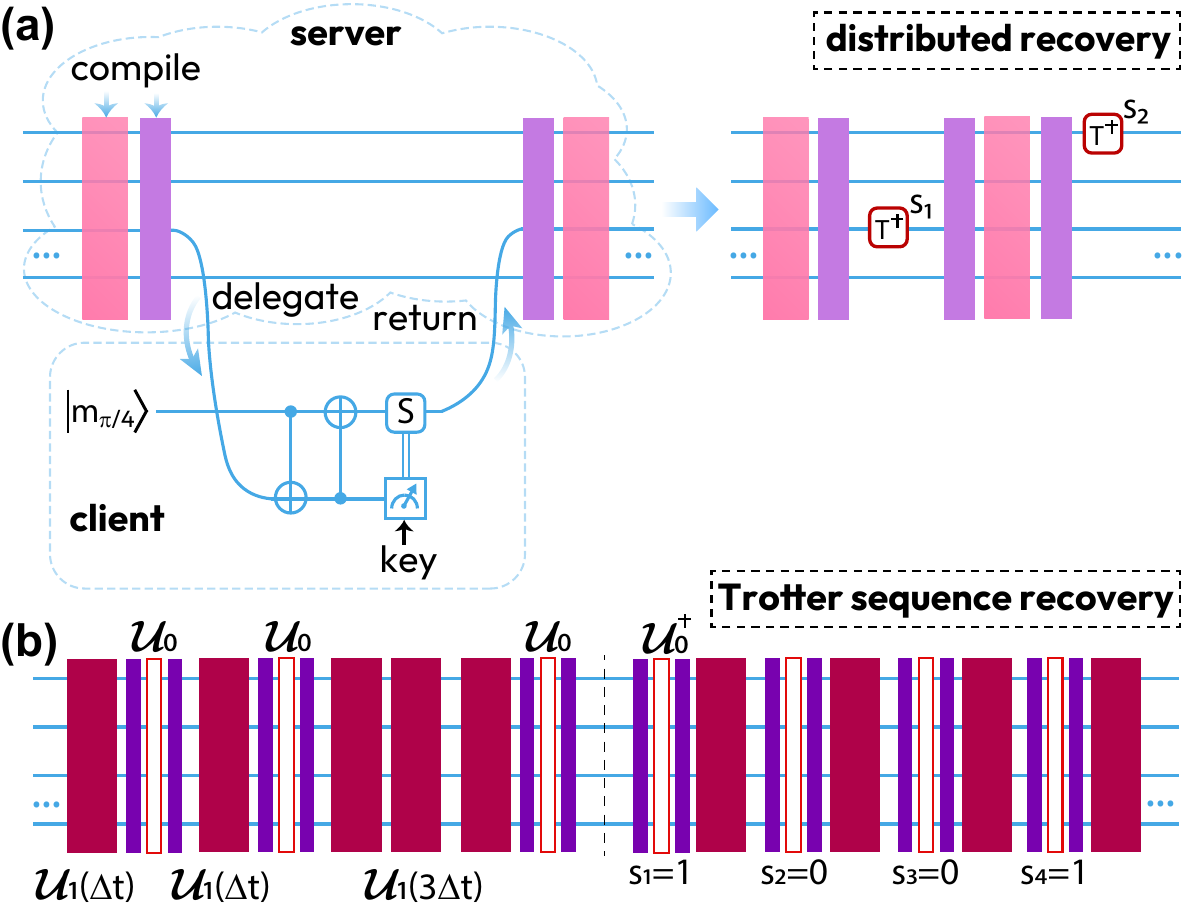}
    \caption{\textbf{Applications of circuit recompilation. (a)} Distributed quantum computing with encrypted gate placement on the client side and server-based compilation. Adaptivity allows the recovery of keys that are lost. \textbf{(b)} Digital quantum evolution (Trotter or QAOA sequence) that is recovered via a discrete step ansatz, where bitstring $s$ allows the matching of varying Trotter steps.}
    \label{fig:applications}
\end{figure}

Let us highlight potential applications of the described optimization beyond showcasing the power of adaptivity. While the setting we studied is rather specific, the core argument---possibility of optimizing circuits for favorable landscapes---remains valid across a wide range of scenarios. One direction lies in circuit recompilation in the regime of \emph{distributed} fault-tolerant quantum computing \cite{DANOS2007,Morimae2012,Sano2022,Polacchi2023,Simmons2024,Main2025}. Non-Clifford T gates are typically implemented through magic state distillation \cite{Bravyi2005,nielsen2010quantum,SalesRodriguez2025} followed by a process of gate teleportation~\cite{ZhouChang2000}, which naturally fits the blind quantum computing mode \cite{Main2025,baranes2025ftqcblind,poshtvan2025selectivelyblind}. Similarly, partially fault-tolerant architectures \cite{Akahoshi2024,Toshio2025} that allow the implementation of arbitrary rotation gates rely on teleportation conditioned on measurements, following repeat-until-success strategies \cite{Bocharov2015}. In each case, some parts of magic generation can be separated from the main compiler and hidden from a server [see Fig.~\ref{fig:applications}(a)]. 
%%%
%%%

Specifically, we consider the case where magic states $|m_{\pi/4}\rangle = (e^{-i\pi/8}|0\rangle + e^{i\pi/8}|0\rangle)/\sqrt{2}$ and $Z$-base measurements \cite{Akahoshi2024} are operated by a client via some cryptographic key $K_C$. This key controls the sequence of generated gates, and the tape $\{T_{q[i]}^{s_i}\}_{i=1}^D$ is the same as long as $K_C$ remains valid. The server sees instructions for qubits being delegated, but does not know measurement outcomes and instructions for placing gates (since these are encrypted). Once the full circuit is compiled in the form similar to Eq.~\eqref{eq:Us}, the resulting state cannot be studied via tomographic techniques. Here, running the server in the adaptive mode can enable effective gate placement learning and potentially lead to the key recovery.

Another example corresponds to circuit recompilation for Trotterized dynamics \cite{Zhuk2024} and similar-looking gate sequences. Consider a multi-qubit Hamiltonian $\mathcal{H} := \mathcal{H}_0 + \mathcal{H}_1$, where $\mathcal{H}_1$ represents some complex (efficiently block-encoded) interaction Hamiltonian, such that $\mathcal{U}_1(t) = \exp(-i\mathcal{H}_1 t)$ leads to scrambling for some nonzero $t$. The evolution with $\mathcal{H}_0$ can be performed efficiently in some entangled basis, $\mathcal{U}_0(t) = V\exp(-iG_0 t)V^\dagger$, where transformation $V$ and generator $G_0$ are provided. We know that the full sequence corresponds to digitized dynamics (potentially, with time-dependent change of terms) or a QAOA-type sequence \cite{ZhuEconomou2022,Chandarana2022}, $\mathcal{U}(\bm{l}^*) := \prod_i \Big[ \mathcal{U}_0(\Delta t) \mathcal{U}_1(l_i \Delta t) \Big]$, with the step size being an integer of $\Delta t$ [Fig.~\ref{fig:applications}(b)]. To understand how much the scrambling unitaries are stretched at each ``bang-bang'' step we can employ an ansatz $\bar{\mathcal{U}}(s) := \prod_i \Big[\mathcal{U}_1^\dagger(\Delta t) \mathcal{U}_0^\dagger(s_i \Delta t)\Big]$ for binary values of $s_i$, where the suitable configuration $s$ can be readily converted into the desired sequence, $s \mapsto \bm{l}$. This Trotter sequence discovery is hence similar to the task we considered before, albeit with T gates substituted by other unitaries. Here, the main feature of the proposed discrete optimization (notable decrease of loss for correctly guessed circuit pattern) shall remain.

\section{Results for a different choice of loss function}\label{sec:diff_loss}

In the main text we have presented analysis for the choice of loss based on the all-zeros projector. While being a faithful loss that works particularly well close to the optimum, other loss functions can provide a good guidance for convergence of recompilation, especially farther away from the solution. One option corresponds to studying a global $Z$-parity operator as $O = Z_1 Z_2 \cdots Z_n = \prod_{j=1}^n Z_j$, and its variance as a measure of the recompilation success. Other options are possible, including using selected strings, pairs, or even just a single-qubit $Z$ operator, leading to quantitative but not qualitative differences. The variational loss thus reads
\begin{equation}
\ell(s) := \mathrm{var}_{| \psi(s) \rangle}[O] = 1 - \langle \psi(s) | O | \psi(s) \rangle^2,
\label{eq:loss_s_parity}
\end{equation}
simplified due to the involutory property of $O^2 = \mathds{1}$. The variance in Eq.~\eqref{eq:loss_s_parity} is bounded between zero and one, $0 \leq \ell(s) \leq 1$, and is clearly minimized at $\ell(s^*) = 0$. Essentially, $\bar{U}(s) \cdot U(s^*) = \mathds{1}$ for $s = s^*$ once the variational bitstring exactly matches the correct placement. For incorrect placements we get $\ell(s) \approx 1$ due to destructive interference between $V_i$-dressed T gates and uncontracted unitaries $W_i$ that pile up into a random sequence. Note that the loss is not faithful as there may be other product states that are prepared by an incorrect puzzle, but the probability for this is exponentially small in the system size, and vanishes on average. 

\begin{figure*}[t!]
\includegraphics[width=1.0\linewidth]{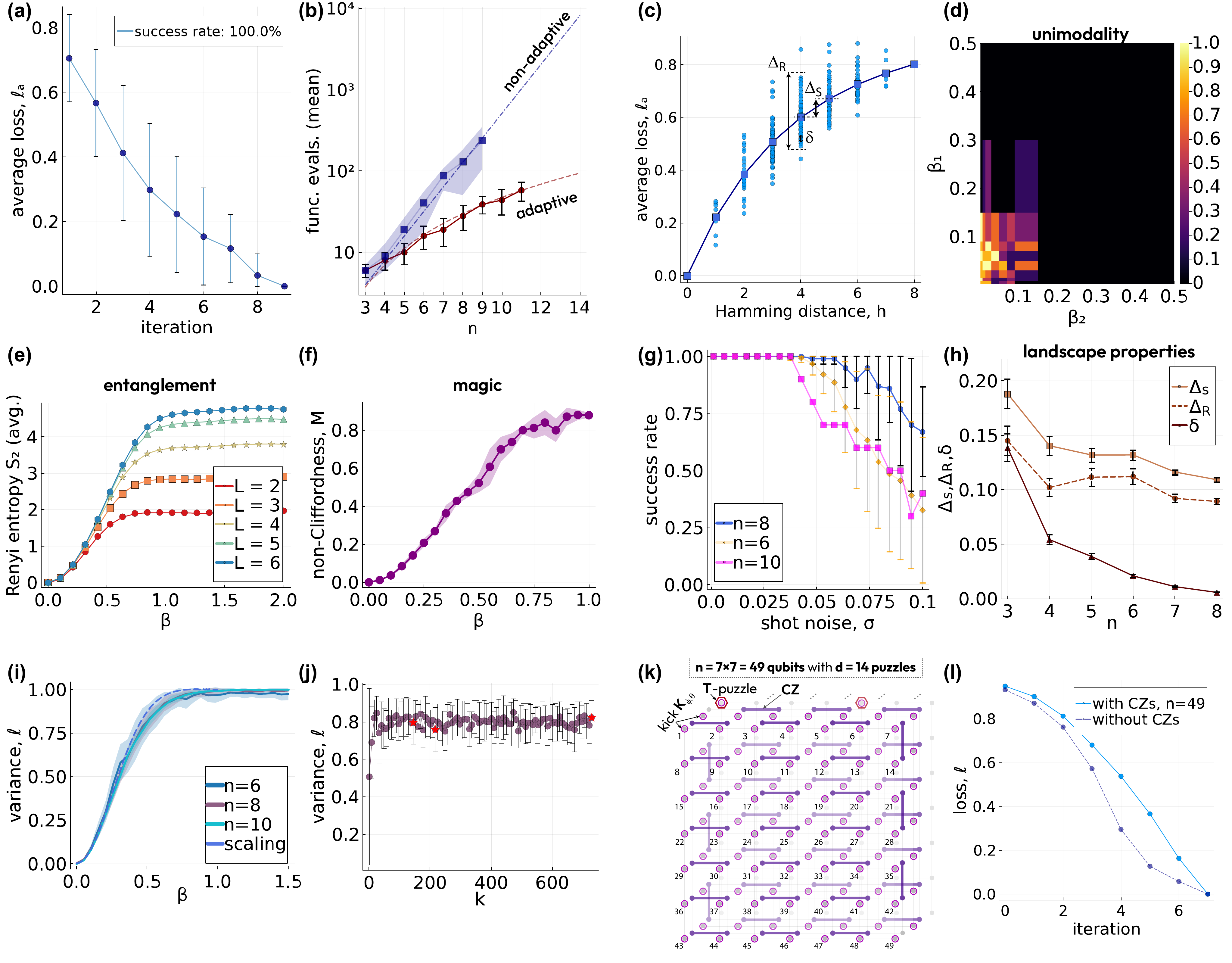}
    \caption{\textbf{Properties of circuit recompilation for modified loss function (global Z parity).} In panels \textbf{(a)}-\textbf{(l)} we show information that corresponds to the same parameters and properties studied in the main text, but with $\ell$ defined as the variance of the $\prod_j Z_j$ operator. \textbf{(a)} Discrete optimization for $n=D=10$ showing perfect success rate within limited number of states. \textbf{(b)} Scaling for discrete optimization (hill climbing) and non-adaptive (random choice) methods at increasing $n=D$. \textbf{(c)} Loss landscape for different optimization bitstrings grouped over Hamming distances to solution. \textbf{(d)} Heatmap of unimodality for $n=D=8$ that show $\beta_{W,V}>0.2$ is sufficient to ensure convergence. \textbf{(e)} Entanglement properties for the effective $\beta$, showing that quantum data we work with have volume law properties. \textbf{(f)} Confirming that circuits have magic that grow with total $\beta_{\mathrm{eff}}$. \textbf{(g)} Noisy hill climbing performed for global parity measurement where shot noise $\sigma$ is finite and larger than characteristic gaps. \textbf{(h)} Three different properties of the landscape, showing that the difference of means has a weak dependence on $n=D$, but gaps for the same Hamming distance decrease with $n$. \textbf{(i)} Variance of the global Z parity as a function of depth of the full circuit (not individual blocks). \textbf{(j)} Dependence of partially-random circuits on the number of terms, for global Z parity. \textbf{(k)} Same as Fig.~6(a) in the main text. \textbf{(l)} Discrete optimization for $n=49$ and $D=14$ showing convergence of hill climbing for the modified loss.}
    \label{fig:parity}
\end{figure*}

The results for the modified global parity loss are presented in Fig.~\ref{fig:parity} (next page), closely related to results in the main text and plotted for the same hyperparameters, system sizes, and number of trials.

\newpage
%%%
%%%

\end{document}

%% file: main.bbl
%apsrev4-2.bst 2019-01-14 (MD) hand-edited version of apsrev4-1.bst
%Control: key (0)
%Control: author (72) initials jnrlst
%Control: editor formatted (1) identically to author
%Control: production of article title (-1) disabled
%Control: page (0) single
%Control: year (1) truncated
%Control: production of eprint (0) enabled
%